\documentclass[longauth]{aa}

\usepackage{graphicx}
\usepackage{natbib}
\usepackage{scalerel}
\usepackage{lastpage}
\usepackage[dvipsnames,table]{xcolor}
\usepackage{comment}
\usepackage{multirow}
\usepackage{booktabs}
\usepackage{soul}
\usepackage{txfonts}
\usepackage{sidecap}

\usepackage[pdfencoding=auto,psdextra]{hyperref}
\hypersetup{
    colorlinks=true,
    linkcolor=blue,
    filecolor=magenta,      
    urlcolor=blue,
    citecolor=blue
}
\makeatletter
\renewcommand*\aa@pageof{, page \thepage{} of \pageref*{LastPage}}
\makeatother

\usepackage[utf8]{inputenc}
\renewcommand{\linenumbers}[0]{}
\usepackage{euclid}

\usepackage[nameinlink,capitalise]{cleveref}

\crefname{section}{Sect.}{Sects.}
\Crefname{section}{Section}{Sections}
\crefname{figure}{Fig.}{Figs.}
\Crefname{figure}{Figure}{Figures}
\crefname{equation}{Eq.}{Eqs.}
\Crefname{equation}{Equation}{Equations}
\crefname{table}{Table}{Tables}
\crefname{appendix}{Appendix}{Appendices}

\usepackage[nolist,nohyperlinks]{acronym}
\newcommand*{\sqdeg}{deg$^{\,2}$\xspace}
\usepackage{orcidlink}

\begin{document}

\title{Euclid Data Release 1 (DR1)}
\subtitle{Persistence in NISP detectors}

\newcommand{\orcid}[1]{} 
\author{Euclid Collaboration: B.~Kubik\orcid{0009-0006-5823-4880}\thanks{\email{bkubik@ipnl.in2p3.fr}}\inst{\ref{aff1}}
\and S.~Conseil\orcid{0000-0002-3657-4191}\inst{\ref{aff2},\ref{aff1}}
\and R.~Barbier\inst{\ref{aff1}}
\and S.~Ferriol\inst{\ref{aff1}}
\and C.~Macabiau\inst{\ref{aff1}}
\and R.~Kohley\inst{\ref{aff3}}
\and P.~Awad\orcid{0000-0002-0428-849X}\inst{\ref{aff4}}
\and A.~Bonchi\orcid{0000-0002-2667-5482}\inst{\ref{aff5}}
\and L.~Bramante\inst{\ref{aff6}}
\and A.~Cillis\orcid{0009-0009-1006-604X}\inst{\ref{aff7},\ref{aff8},\ref{aff9}}
\and M.~Correnti\orcid{0000-0001-6464-3257}\inst{\ref{aff10},\ref{aff5}}
\and R.~da~Silva\orcid{0000-0003-4788-677X}\inst{\ref{aff10},\ref{aff5}}
\and I.~Das\orcid{0009-0007-7088-2044}\inst{\ref{aff11}}
\and F.~Faustini\orcid{0000-0001-6274-5145}\inst{\ref{aff10}}
\and R.~Foltz\inst{\ref{aff8}}
\and M.~Frailis\orcid{0000-0002-7400-2135}\inst{\ref{aff12}}
\and Y.~Fu\orcid{0000-0002-0759-0504}\inst{\ref{aff4},\ref{aff13}}
\and T.~Gasparetto\orcid{0000-0002-7913-4866}\inst{\ref{aff10}}
\and W.~Gillard\orcid{0000-0003-4744-9748}\inst{\ref{aff14}}
\and S.~Hemmati\orcid{0000-0003-2226-5395}\inst{\ref{aff11}}
\and J.~Jacobson\inst{\ref{aff11}}
\and K.~Jahnke\orcid{0000-0003-3804-2137}\inst{\ref{aff15}}
\and P.~W.~Morris\orcid{0000-0002-5186-4381}\inst{\ref{aff16}}
\and K.~Paterson\orcid{0000-0001-8340-3486}\inst{\ref{aff15}}
\and G.~Polenta\orcid{0000-0003-4067-9196}\inst{\ref{aff5}}
\and M.~Radovich\orcid{0000-0002-3585-866X}\inst{\ref{aff17}}
\and M.~Schirmer\orcid{0000-0003-2568-9994}\inst{\ref{aff15}}
\and H.~I.~Teplitz\orcid{0000-0002-7064-5424}\inst{\ref{aff18}}
\and A.~Troja\orcid{0000-0003-0239-4595}\inst{\ref{aff12}}
\and B.~P.~Venemans\orcid{0000-0001-9024-8322}\inst{\ref{aff4}}
\and B.~Altieri\orcid{0000-0003-3936-0284}\inst{\ref{aff3}}
\and S.~Andreon\orcid{0000-0002-2041-8784}\inst{\ref{aff19}}
\and N.~Auricchio\orcid{0000-0003-4444-8651}\inst{\ref{aff20}}
\and H.~Aussel\orcid{0000-0002-1371-5705}\inst{\ref{aff21}}
\and C.~Baccigalupi\orcid{0000-0002-8211-1630}\inst{\ref{aff22},\ref{aff12},\ref{aff23},\ref{aff24}}
\and M.~Baldi\orcid{0000-0003-4145-1943}\inst{\ref{aff25},\ref{aff20},\ref{aff26}}
\and A.~Balestra\orcid{0000-0002-6967-261X}\inst{\ref{aff17}}
\and S.~Bardelli\orcid{0000-0002-8900-0298}\inst{\ref{aff20}}
\and P.~Battaglia\orcid{0000-0002-7337-5909}\inst{\ref{aff20}}
\and A.~Biviano\orcid{0000-0002-0857-0732}\inst{\ref{aff12},\ref{aff22}}
\and E.~Branchini\orcid{0000-0002-0808-6908}\inst{\ref{aff27},\ref{aff28},\ref{aff19}}
\and M.~Brescia\orcid{0000-0001-9506-5680}\inst{\ref{aff29},\ref{aff30}}
\and J.~Brinchmann\orcid{0000-0003-4359-8797}\inst{\ref{aff31},\ref{aff32},\ref{aff33}}
\and S.~Camera\orcid{0000-0003-3399-3574}\inst{\ref{aff34},\ref{aff35},\ref{aff36}}
\and G.~Ca\~nas-Herrera\orcid{0000-0003-2796-2149}\inst{\ref{aff4}}
\and V.~Capobianco\orcid{0000-0002-3309-7692}\inst{\ref{aff36}}
\and C.~Carbone\orcid{0000-0003-0125-3563}\inst{\ref{aff37}}
\and J.~Carretero\orcid{0000-0002-3130-0204}\inst{\ref{aff38},\ref{aff39}}
\and F.~J.~Castander\orcid{0000-0001-7316-4573}\inst{\ref{aff40},\ref{aff41}}
\and M.~Castellano\orcid{0000-0001-9875-8263}\inst{\ref{aff10}}
\and G.~Castignani\orcid{0000-0001-6831-0687}\inst{\ref{aff20}}
\and S.~Cavuoti\orcid{0000-0002-3787-4196}\inst{\ref{aff30},\ref{aff42}}
\and K.~C.~Chambers\orcid{0000-0001-6965-7789}\inst{\ref{aff43}}
\and A.~Cimatti\inst{\ref{aff44}}
\and C.~Colodro-Conde\inst{\ref{aff45}}
\and G.~Congedo\orcid{0000-0003-2508-0046}\inst{\ref{aff46}}
\and C.~J.~Conselice\orcid{0000-0003-1949-7638}\inst{\ref{aff47}}
\and L.~Conversi\orcid{0000-0002-6710-8476}\inst{\ref{aff48},\ref{aff3}}
\and Y.~Copin\orcid{0000-0002-5317-7518}\inst{\ref{aff1}}
\and F.~Courbin\orcid{0000-0003-0758-6510}\inst{\ref{aff49},\ref{aff50},\ref{aff51}}
\and H.~M.~Courtois\orcid{0000-0003-0509-1776}\inst{\ref{aff52}}
\and A.~Da~Silva\orcid{0000-0002-6385-1609}\inst{\ref{aff53},\ref{aff54}}
\and H.~Degaudenzi\orcid{0000-0002-5887-6799}\inst{\ref{aff55}}
\and S.~de~la~Torre\inst{\ref{aff56}}
\and G.~De~Lucia\orcid{0000-0002-6220-9104}\inst{\ref{aff12}}
\and H.~Dole\orcid{0000-0002-9767-3839}\inst{\ref{aff57}}
\and F.~Dubath\orcid{0000-0002-6533-2810}\inst{\ref{aff55}}
\and X.~Dupac\inst{\ref{aff3}}
\and S.~Dusini\orcid{0000-0002-1128-0664}\inst{\ref{aff58}}
\and S.~Escoffier\orcid{0000-0002-2847-7498}\inst{\ref{aff14}}
\and M.~Farina\orcid{0000-0002-3089-7846}\inst{\ref{aff59}}
\and R.~Farinelli\inst{\ref{aff20}}
\and S.~Fotopoulou\orcid{0000-0002-9686-254X}\inst{\ref{aff60}}
\and N.~Fourmanoit\orcid{0009-0005-6816-6925}\inst{\ref{aff14}}
\and E.~Franceschi\orcid{0000-0002-0585-6591}\inst{\ref{aff20}}
\and M.~Fumana\orcid{0000-0001-6787-5950}\inst{\ref{aff37}}
\and L.~Gabarra\orcid{0000-0002-8486-8856}\inst{\ref{aff61}}
\and S.~Galeotta\orcid{0000-0002-3748-5115}\inst{\ref{aff12}}
\and K.~George\orcid{0000-0002-1734-8455}\inst{\ref{aff62}}
\and B.~Gillis\orcid{0000-0002-4478-1270}\inst{\ref{aff46}}
\and C.~Giocoli\orcid{0000-0002-9590-7961}\inst{\ref{aff20},\ref{aff26}}
\and J.~Gracia-Carpio\orcid{0000-0003-4689-3134}\inst{\ref{aff63}}
\and B.~R.~Granett\orcid{0000-0003-2694-9284}\inst{\ref{aff19}}
\and A.~Grazian\orcid{0000-0002-5688-0663}\inst{\ref{aff17}}
\and F.~Grupp\inst{\ref{aff63},\ref{aff64}}
\and L.~Guzzo\orcid{0000-0001-8264-5192}\inst{\ref{aff65},\ref{aff19},\ref{aff66}}
\and W.~G.~Hartley\orcid{0000-0002-4563-5101}\inst{\ref{aff55}}
\and S.~V.~H.~Haugan\orcid{0000-0001-9648-7260}\inst{\ref{aff67}}
\and J.~Hoar\inst{\ref{aff3}}
\and H.~Hoekstra\orcid{0000-0002-0641-3231}\inst{\ref{aff4}}
\and W.~Holmes\orcid{0009-0007-8554-4646}\inst{\ref{aff68}}
\and F.~Hormuth\inst{\ref{aff69}}
\and A.~Hornstrup\orcid{0000-0002-3363-0936}\inst{\ref{aff70},\ref{aff71}}
\and P.~Hudelot\inst{\ref{aff72}}
\and M.~Jhabvala\inst{\ref{aff8}}
\and S.~Kermiche\orcid{0000-0002-0302-5735}\inst{\ref{aff14}}
\and A.~Kiessling\orcid{0000-0002-2590-1273}\inst{\ref{aff68}}
\and M.~Kunz\orcid{0000-0002-3052-7394}\inst{\ref{aff73}}
\and H.~Kurki-Suonio\orcid{0000-0002-4618-3063}\inst{\ref{aff74},\ref{aff75}}
\and A.~M.~C.~Le~Brun\orcid{0000-0002-0936-4594}\inst{\ref{aff76}}
\and V.~Le~Brun\orcid{0000-0002-5027-1939}\inst{\ref{aff56}}
\and D.~Le~Mignant\orcid{0000-0002-5339-5515}\inst{\ref{aff56}}
\and P.~Liebing\orcid{0009-0005-8043-7150}\inst{\ref{aff77}}
\and S.~Ligori\orcid{0000-0003-4172-4606}\inst{\ref{aff36}}
\and P.~B.~Lilje\orcid{0000-0003-4324-7794}\inst{\ref{aff67}}
\and V.~Lindholm\orcid{0000-0003-2317-5471}\inst{\ref{aff74},\ref{aff75}}
\and I.~Lloro\orcid{0000-0001-5966-1434}\inst{\ref{aff78}}
\and M.~Magliocchetti\orcid{0000-0001-9158-4838}\inst{\ref{aff59}}
\and G.~Mainetti\orcid{0000-0003-2384-2377}\inst{\ref{aff79}}
\and D.~Maino\inst{\ref{aff65},\ref{aff37},\ref{aff66}}
\and E.~Maiorano\orcid{0000-0003-2593-4355}\inst{\ref{aff20}}
\and O.~Mansutti\orcid{0000-0001-5758-4658}\inst{\ref{aff12}}
\and S.~Marcin\inst{\ref{aff80}}
\and O.~Marggraf\orcid{0000-0001-7242-3852}\inst{\ref{aff81}}
\and K.~Markovic\orcid{0000-0001-6764-073X}\inst{\ref{aff68}}
\and M.~Martinelli\orcid{0000-0002-6943-7732}\inst{\ref{aff10},\ref{aff82}}
\and N.~Martinet\orcid{0000-0003-2786-7790}\inst{\ref{aff56}}
\and F.~Marulli\orcid{0000-0002-8850-0303}\inst{\ref{aff83},\ref{aff20},\ref{aff26}}
\and R.~J.~Massey\orcid{0000-0002-6085-3780}\inst{\ref{aff84}}
\and N.~Mauri\orcid{0000-0001-8196-1548}\inst{\ref{aff44},\ref{aff26}}
\and C.~J.~R.~McPartland\orcid{0000-0003-0639-025X}\inst{\ref{aff71},\ref{aff85}}
\and E.~Medinaceli\orcid{0000-0002-4040-7783}\inst{\ref{aff20}}
\and S.~Mei\orcid{0000-0002-2849-559X}\inst{\ref{aff86},\ref{aff87}}
\and M.~Meneghetti\orcid{0000-0003-1225-7084}\inst{\ref{aff20},\ref{aff26}}
\and E.~Merlin\orcid{0000-0001-6870-8900}\inst{\ref{aff17}}
\and G.~Meylan\orcid{0000-0001-6503-0209}\inst{\ref{aff88}}
\and P.~Monaco\orcid{0000-0003-2083-7564}\inst{\ref{aff89},\ref{aff12},\ref{aff23},\ref{aff22}}
\and A.~Mora\orcid{0000-0002-1922-8529}\inst{\ref{aff90}}
\and M.~Moresco\orcid{0000-0002-7616-7136}\inst{\ref{aff83},\ref{aff20}}
\and C.~Moretti\orcid{0000-0003-3314-8936}\inst{\ref{aff12},\ref{aff22},\ref{aff23}}
\and L.~Moscardini\orcid{0000-0002-3473-6716}\inst{\ref{aff83},\ref{aff20},\ref{aff26}}
\and C.~Neissner\orcid{0000-0001-8524-4968}\inst{\ref{aff91},\ref{aff39}}
\and R.~C.~Nichol\orcid{0000-0003-0939-6518}\inst{\ref{aff92}}
\and S.-M.~Niemi\orcid{0009-0005-0247-0086}\inst{\ref{aff93}}
\and C.~Padilla\orcid{0000-0001-7951-0166}\inst{\ref{aff91}}
\and S.~Paltani\orcid{0000-0002-8108-9179}\inst{\ref{aff55}}
\and F.~Pasian\orcid{0000-0002-4869-3227}\inst{\ref{aff12}}
\and W.~J.~Percival\orcid{0000-0002-0644-5727}\inst{\ref{aff94},\ref{aff95},\ref{aff96}}
\and V.~Pettorino\orcid{0000-0002-4203-9320}\inst{\ref{aff93}}
\and A.~Pezzotta\orcid{0000-0003-0726-2268}\inst{\ref{aff19}}
\and S.~Pires\orcid{0000-0002-0249-2104}\inst{\ref{aff21}}
\and L.~A.~Popa\inst{\ref{aff97}}
\and L.~Pozzetti\orcid{0000-0001-7085-0412}\inst{\ref{aff20}}
\and G.~D.~Racca\orcid{0000-0002-9883-8981}\inst{\ref{aff4},\ref{aff93}}
\and F.~Raison\orcid{0000-0002-7819-6918}\inst{\ref{aff63}}
\and R.~Rebolo\orcid{0000-0003-3767-7085}\inst{\ref{aff45},\ref{aff98},\ref{aff99}}
\and A.~Refregier\inst{\ref{aff100}}
\and A.~Renzi\orcid{0000-0001-9856-1970}\inst{\ref{aff101},\ref{aff58},\ref{aff20}}
\and J.~Rhodes\orcid{0000-0002-4485-8549}\inst{\ref{aff68}}
\and G.~Riccio\inst{\ref{aff30}}
\and I.~Risso\orcid{0000-0003-2525-7761}\inst{\ref{aff27},\ref{aff28},\ref{aff19}}
\and F.~Rizzo\orcid{0000-0002-9407-585X}\inst{\ref{aff12}}
\and E.~Romelli\orcid{0000-0003-3069-9222}\inst{\ref{aff12}}
\and M.~Roncarelli\orcid{0000-0001-9587-7822}\inst{\ref{aff20}}
\and C.~Rosset\orcid{0000-0003-0286-2192}\inst{\ref{aff86}}
\and B.~Rusholme\orcid{0000-0001-7648-4142}\inst{\ref{aff11}}
\and R.~Saglia\orcid{0000-0003-0378-7032}\inst{\ref{aff64},\ref{aff63}}
\and Z.~Sakr\orcid{0000-0002-4823-3757}\inst{\ref{aff102},\ref{aff103},\ref{aff104}}
\and A.~G.~S\'anchez\orcid{0000-0003-1198-831X}\inst{\ref{aff63}}
\and D.~Sapone\orcid{0000-0001-7089-4503}\inst{\ref{aff105}}
\and B.~Sartoris\orcid{0000-0003-1337-5269}\inst{\ref{aff63},\ref{aff12}}
\and P.~Schneider\orcid{0000-0001-8561-2679}\inst{\ref{aff81}}
\and M.~Scodeggio\inst{\ref{aff37}}
\and A.~Secroun\orcid{0000-0003-0505-3710}\inst{\ref{aff14}}
\and M.~Seiffert\orcid{0000-0002-7536-9393}\inst{\ref{aff68}}
\and E.~Sihvola\orcid{0000-0003-1804-7715}\inst{\ref{aff106}}
\and C.~Sirignano\orcid{0000-0002-0995-7146}\inst{\ref{aff101},\ref{aff58}}
\and G.~Sirri\orcid{0000-0003-2626-2853}\inst{\ref{aff26}}
\and L.~Stanco\orcid{0000-0002-9706-5104}\inst{\ref{aff58}}
\and P.~Tallada-Cresp\'{i}\orcid{0000-0002-1336-8328}\inst{\ref{aff38},\ref{aff39}}
\and A.~N.~Taylor\inst{\ref{aff46}}
\and I.~Tereno\orcid{0000-0002-4537-6218}\inst{\ref{aff53},\ref{aff107}}
\and N.~Tessore\orcid{0000-0002-9696-7931}\inst{\ref{aff77}}
\and S.~Toft\orcid{0000-0003-3631-7176}\inst{\ref{aff108},\ref{aff85}}
\and R.~Toledo-Moreo\orcid{0000-0002-2997-4859}\inst{\ref{aff109},\ref{aff110}}
\and F.~Torradeflot\orcid{0000-0003-1160-1517}\inst{\ref{aff39},\ref{aff38}}
\and I.~Tutusaus\orcid{0000-0002-3199-0399}\inst{\ref{aff40},\ref{aff41},\ref{aff103}}
\and J.~Valiviita\orcid{0000-0001-6225-3693}\inst{\ref{aff74},\ref{aff75}}
\and T.~Vassallo\orcid{0000-0001-6512-6358}\inst{\ref{aff12},\ref{aff62}}
\and A.~Veropalumbo\orcid{0000-0003-2387-1194}\inst{\ref{aff19},\ref{aff28},\ref{aff27}}
\and S.~Wachter\orcid{0009-0007-5938-3199}\inst{\ref{aff111}}
\and Y.~Wang\orcid{0000-0002-4749-2984}\inst{\ref{aff11}}
\and J.~Weller\orcid{0000-0002-8282-2010}\inst{\ref{aff64},\ref{aff63}}
\and A.~Zacchei\orcid{0000-0003-0396-1192}\inst{\ref{aff12},\ref{aff22}}
\and G.~Zamorani\orcid{0000-0002-2318-301X}\inst{\ref{aff20}}
\and E.~Zucca\orcid{0000-0002-5845-8132}\inst{\ref{aff20}}
\and M.~Ballardini\orcid{0000-0003-4481-3559}\inst{\ref{aff112},\ref{aff113},\ref{aff20}}
\and P.~Bergamini\orcid{0000-0003-1383-9414}\inst{\ref{aff20}}
\and S.~Borgani\orcid{0000-0001-6151-6439}\inst{\ref{aff89},\ref{aff22},\ref{aff12},\ref{aff23},\ref{aff114}}
\and E.~Bozzo\orcid{0000-0002-8201-1525}\inst{\ref{aff55}}
\and C.~Burigana\orcid{0000-0002-3005-5796}\inst{\ref{aff115},\ref{aff116}}
\and R.~Cabanac\orcid{0000-0001-6679-2600}\inst{\ref{aff103}}
\and M.~Calabrese\orcid{0000-0002-2637-2422}\inst{\ref{aff117},\ref{aff37}}
\and F.~Caro\orcid{0009-0003-1053-0507}\inst{\ref{aff10}}
\and T.~Castro\orcid{0000-0002-6292-3228}\inst{\ref{aff118},\ref{aff12},\ref{aff22}}
\and D.~Di~Ferdinando\inst{\ref{aff26}}
\and J.~A.~Escartin~Vigo\inst{\ref{aff63}}
\and G.~Fabbian\orcid{0000-0002-3255-4695}\inst{\ref{aff57}}
\and J.~Garc\'ia-Bellido\orcid{0000-0002-9370-8360}\inst{\ref{aff102}}
\and V.~Gautard\inst{\ref{aff119}}
\and A.~Gregorio\orcid{0000-0003-4028-8785}\inst{\ref{aff89},\ref{aff12},\ref{aff23}}
\and C.~M.~Gutierrez\orcid{0000-0001-7854-783X}\inst{\ref{aff45},\ref{aff99}}
\and A.~Kashlinsky\orcid{0000-0003-2156-078X}\inst{\ref{aff120},\ref{aff121},\ref{aff122}}
\and A.~Loureiro\orcid{0000-0002-4371-0876}\inst{\ref{aff123},\ref{aff124}}
\and J.~Macias-Perez\orcid{0000-0002-5385-2763}\inst{\ref{aff125}}
\and R.~Maoli\orcid{0000-0002-6065-3025}\inst{\ref{aff126},\ref{aff10}}
\and R.~B.~Metcalf\orcid{0000-0003-3167-2574}\inst{\ref{aff83},\ref{aff20}}
\and A.~A.~Nucita\inst{\ref{aff127},\ref{aff128},\ref{aff129}}
\and M.~P\"ontinen\orcid{0000-0001-5442-2530}\inst{\ref{aff74}}
\and E.~Sarpa\orcid{0000-0002-1256-655X}\inst{\ref{aff12}}
\and V.~Scottez\orcid{0009-0008-3864-940X}\inst{\ref{aff130},\ref{aff131}}
\and M.~Sereno\orcid{0000-0003-0302-0325}\inst{\ref{aff20},\ref{aff26}}
\and M.~Tenti\orcid{0000-0002-4254-5901}\inst{\ref{aff26}}
\and M.~Tucci\inst{\ref{aff55}}
\and M.~Viel\orcid{0000-0002-2642-5707}\inst{\ref{aff22},\ref{aff12},\ref{aff24},\ref{aff23},\ref{aff114}}
\and M.~Wiesmann\orcid{0009-0000-8199-5860}\inst{\ref{aff67}}
\and J.~A.~Acevedo~Barroso\orcid{0000-0002-9654-1711}\inst{\ref{aff68}}
\and Y.~Akrami\orcid{0000-0002-2407-7956}\inst{\ref{aff102},\ref{aff132}}
\and I.~T.~Andika\orcid{0000-0001-6102-9526}\inst{\ref{aff64}}
\and G.~Angora\orcid{0000-0002-0316-6562}\inst{\ref{aff30},\ref{aff112}}
\and S.~Anselmi\orcid{0000-0002-3579-9583}\inst{\ref{aff58},\ref{aff101},\ref{aff133}}
\and M.~Archidiacono\orcid{0000-0003-4952-9012}\inst{\ref{aff65},\ref{aff66}}
\and G.~Aric\`o\orcid{0000-0002-2802-2928}\inst{\ref{aff26}}
\and F.~Atrio-Barandela\orcid{0000-0002-2130-2513}\inst{\ref{aff134}}
\and M.~Baes\orcid{0000-0002-3930-2757}\inst{\ref{aff135}}
\and L.~Bazzanini\orcid{0000-0003-0727-0137}\inst{\ref{aff112},\ref{aff20}}
\and D.~Bertacca\orcid{0000-0002-2490-7139}\inst{\ref{aff101},\ref{aff17},\ref{aff58}}
\and M.~Bethermin\orcid{0000-0002-3915-2015}\inst{\ref{aff136}}
\and F.~Beutler\orcid{0000-0003-0467-5438}\inst{\ref{aff46}}
\and L.~Blot\orcid{0000-0002-9622-7167}\inst{\ref{aff137},\ref{aff76}}
\and M.~L.~Brown\orcid{0000-0002-0370-8077}\inst{\ref{aff47}}
\and S.~Bruton\orcid{0000-0002-6503-5218}\inst{\ref{aff16}}
\and A.~Calabro\orcid{0000-0003-2536-1614}\inst{\ref{aff10}}
\and B.~Camacho~Quevedo\orcid{0000-0002-8789-4232}\inst{\ref{aff22},\ref{aff24},\ref{aff12}}
\and C.~S.~Carvalho\inst{\ref{aff107}}
\and Y.~Charles\orcid{0009-0000-3636-834X}\inst{\ref{aff56}}
\and F.~Cogato\orcid{0000-0003-4632-6113}\inst{\ref{aff83},\ref{aff20}}
\and T.~E.~Collett\orcid{0000-0001-5564-3140}\inst{\ref{aff138}}
\and A.~R.~Cooray\orcid{0000-0002-3892-0190}\inst{\ref{aff139}}
\and P.~Corcho-Caballero\orcid{0000-0001-6327-7080}\inst{\ref{aff13}}
\and B.~Csizi\orcid{0000-0003-3227-6581}\inst{\ref{aff140}}
\and O.~Cucciati\orcid{0000-0002-9336-7551}\inst{\ref{aff20}}
\and J.~E.~Davies\orcid{0000-0002-5079-9098}\inst{\ref{aff15}}
\and S.~Davini\orcid{0000-0003-3269-1718}\inst{\ref{aff28}}
\and T.~de~Boer\orcid{0000-0001-5486-2747}\inst{\ref{aff43}}
\and G.~Desprez\orcid{0000-0001-8325-1742}\inst{\ref{aff13}}
\and A.~D\'iaz-S\'anchez\orcid{0000-0003-0748-4768}\inst{\ref{aff141}}
\and S.~Di~Domizio\orcid{0000-0003-2863-5895}\inst{\ref{aff27},\ref{aff28}}
\and J.~M.~Diego\orcid{0000-0001-9065-3926}\inst{\ref{aff142}}
\and P.-A.~Duc\orcid{0000-0003-3343-6284}\inst{\ref{aff136}}
\and V.~Duret\orcid{0009-0009-0383-4960}\inst{\ref{aff14}}
\and M.~Y.~Elkhashab\orcid{0000-0001-9306-2603}\inst{\ref{aff12},\ref{aff23},\ref{aff89},\ref{aff22}}
\and Y.~Fang\orcid{0000-0002-0334-6950}\inst{\ref{aff64}}
\and A.~Farina\orcid{0009-0000-3420-929X}\inst{\ref{aff19},\ref{aff28}}
\and A.~M.~N.~Ferguson\orcid{0000-0001-7934-1278}\inst{\ref{aff46}}
\and A.~Finoguenov\orcid{0000-0002-4606-5403}\inst{\ref{aff74}}
\and A.~Franco\orcid{0000-0002-4761-366X}\inst{\ref{aff127},\ref{aff128},\ref{aff129}}
\and K.~Ganga\orcid{0000-0001-8159-8208}\inst{\ref{aff86}}
\and E.~Gaztanaga\orcid{0000-0001-9632-0815}\inst{\ref{aff40},\ref{aff41},\ref{aff138}}
\and Z.~Ghaffari\orcid{0000-0002-6467-8078}\inst{\ref{aff12},\ref{aff22}}
\and F.~Giacomini\orcid{0000-0002-3129-2814}\inst{\ref{aff26}}
\and F.~Gianotti\orcid{0000-0003-4666-119X}\inst{\ref{aff20}}
\and A.~H.~Gonzalez\orcid{0000-0002-0933-8601}\inst{\ref{aff143}}
\and G.~Gozaliasl\orcid{0000-0002-0236-919X}\inst{\ref{aff144},\ref{aff74}}
\and A.~Gruppuso\orcid{0000-0001-9272-5292}\inst{\ref{aff20},\ref{aff26}}
\and M.~Guidi\orcid{0000-0001-9408-1101}\inst{\ref{aff25},\ref{aff20}}
\and A.~Hall\orcid{0000-0002-3139-8651}\inst{\ref{aff46}}
\and N.~A.~Hatch\orcid{0000-0001-5600-0534}\inst{\ref{aff145}}
\and T.~Heuberger\orcid{0009-0009-3474-8467}\inst{\ref{aff80}}
\and H.~Hildebrandt\orcid{0000-0002-9814-3338}\inst{\ref{aff146}}
\and J.~Hjorth\orcid{0000-0002-4571-2306}\inst{\ref{aff147}}
\and L.~K.~Hunt\orcid{0000-0001-9162-2371}\inst{\ref{aff148}}
\and M.~Jauzac\orcid{0000-0003-1974-8732}\inst{\ref{aff149},\ref{aff84},\ref{aff150},\ref{aff151},\ref{aff103}}
\and J.~J.~E.~Kajava\orcid{0000-0002-3010-8333}\inst{\ref{aff152},\ref{aff153},\ref{aff154}}
\and Y.~Kang\orcid{0009-0000-8588-7250}\inst{\ref{aff140}}
\and V.~Kansal\orcid{0000-0002-4008-6078}\inst{\ref{aff155},\ref{aff156}}
\and D.~Karagiannis\orcid{0000-0002-4927-0816}\inst{\ref{aff112},\ref{aff157}}
\and J.~Kim\orcid{0000-0003-2776-2761}\inst{\ref{aff61}}
\and C.~C.~Kirkpatrick\inst{\ref{aff106}}
\and A.~Kov\'acs\orcid{0000-0002-5825-579X}\inst{\ref{aff158},\ref{aff159}}
\and I.~Kova{\v{c}}i{\'{c}}\orcid{0000-0001-6751-3263}\inst{\ref{aff135}}
\and K.~Koyama\orcid{0000-0001-6727-6915}\inst{\ref{aff138}}
\and S.~Kruk\orcid{0000-0001-8010-8879}\inst{\ref{aff3}}
\and M.~C.~Lam\orcid{0000-0002-9347-2298}\inst{\ref{aff46}}
\and M.~Lattanzi\orcid{0000-0003-1059-2532}\inst{\ref{aff113}}
\and J.~Le~Graet\orcid{0000-0001-6523-7971}\inst{\ref{aff14}}
\and G.~Leroy\orcid{0009-0004-2523-4425}\inst{\ref{aff149},\ref{aff84}}
\and J.~Lesgourgues\orcid{0000-0001-7627-353X}\inst{\ref{aff160}}
\and T.~I.~Liaudat\orcid{0000-0002-9104-314X}\inst{\ref{aff161}}
\and L.~Linke\orcid{0000-0002-2622-8113}\inst{\ref{aff140}}
\and S.~J.~Liu\orcid{0000-0001-7680-2139}\inst{\ref{aff59}}
\and G.~Maggio\orcid{0000-0003-4020-4836}\inst{\ref{aff12}}
\and F.~Mannucci\orcid{0000-0002-4803-2381}\inst{\ref{aff148}}
\and F.~R.~Marleau\orcid{0000-0002-1442-2947}\inst{\ref{aff140}}
\and C.~J.~A.~P.~Martins\orcid{0000-0002-4886-9261}\inst{\ref{aff162},\ref{aff31}}
\and T.~MENARD\inst{\ref{aff130}}
\and M.~Miluzio\inst{\ref{aff3},\ref{aff163}}
\and G.~Morgante\inst{\ref{aff20}}
\and S.~Nadathur\orcid{0000-0001-9070-3102}\inst{\ref{aff138}}
\and K.~Naidoo\orcid{0000-0002-9182-1802}\inst{\ref{aff138},\ref{aff15}}
\and P.~Natoli\orcid{0000-0003-0126-9100}\inst{\ref{aff112},\ref{aff113}}
\and A.~Navarro-Alsina\orcid{0000-0002-3173-2592}\inst{\ref{aff81}}
\and S.~Nesseris\orcid{0000-0002-0567-0324}\inst{\ref{aff102}}
\and L.~Nicastro\orcid{0000-0001-8534-6788}\inst{\ref{aff20}}
\and F.~Oppizzi\orcid{0000-0003-3904-8370}\inst{\ref{aff28},\ref{aff58},\ref{aff101}}
\and F.~Pace\orcid{0000-0001-8039-0480}\inst{\ref{aff34},\ref{aff35},\ref{aff36},\ref{aff54}}
\and E.~Palazzi\orcid{0000-0002-8691-7666}\inst{\ref{aff20}}
\and D.~Paoletti\orcid{0000-0003-4761-6147}\inst{\ref{aff20},\ref{aff116}}
\and F.~Passalacqua\orcid{0000-0002-8606-4093}\inst{\ref{aff58}}
\and L.~Patrizii\inst{\ref{aff26}}
\and C.~Pattison\orcid{0000-0003-3272-2617}\inst{\ref{aff138}}
\and A.~Pisani\orcid{0000-0002-6146-4437}\inst{\ref{aff14}}
\and D.~Potter\orcid{0000-0002-0757-5195}\inst{\ref{aff164}}
\and G.~W.~Pratt\inst{\ref{aff21}}
\and S.~Quai\orcid{0000-0002-0449-8163}\inst{\ref{aff83},\ref{aff20}}
\and G.~Rodighiero\orcid{0000-0002-9415-2296}\inst{\ref{aff101},\ref{aff17}}
\and W.~Roster\orcid{0000-0002-9149-6528}\inst{\ref{aff63}}
\and M.~Sahl\'en\orcid{0000-0003-0973-4804}\inst{\ref{aff165}}
\and D.~B.~Sanders\orcid{0000-0002-1233-9998}\inst{\ref{aff43}}
\and A.~Schneider\orcid{0000-0001-7055-8104}\inst{\ref{aff164}}
\and D.~Sciotti\orcid{0009-0008-4519-2620}\inst{\ref{aff10},\ref{aff82}}
\and E.~Sellentin\orcid{0009-0002-2655-3458}\inst{\ref{aff166},\ref{aff4}}
\and S.~Serjeant\orcid{0000-0002-0517-7943}\inst{\ref{aff167}}
\and A.~Shulevski\orcid{0000-0002-1827-0469}\inst{\ref{aff168},\ref{aff13},\ref{aff169}}
\and L.~C.~Smith\orcid{0000-0002-3259-2771}\inst{\ref{aff170}}
\and I.~Szapudi\orcid{0000-0003-2274-0301}\inst{\ref{aff43}}
\and M.~Talia\orcid{0000-0003-4352-2063}\inst{\ref{aff83},\ref{aff20}}
\and K.~Tanidis\orcid{0000-0001-9843-5130}\inst{\ref{aff171}}
\and C.~Tao\orcid{0000-0001-7961-8177}\inst{\ref{aff14}}
\and G.~Testera\orcid{0000-0003-2970-766X}\inst{\ref{aff28}}
\and R.~Teyssier\orcid{0000-0001-7689-0933}\inst{\ref{aff172}}
\and S.~Tosi\orcid{0000-0002-7275-9193}\inst{\ref{aff27},\ref{aff19},\ref{aff28}}
\and C.~Uhlemann\orcid{0000-0001-7831-1579}\inst{\ref{aff173},\ref{aff174}}
\and M.~Urbano\orcid{0000-0001-5640-0650}\inst{\ref{aff145},\ref{aff136}}
\and C.~Valieri\inst{\ref{aff26}}
\and A.~Venhola\orcid{0000-0001-6071-4564}\inst{\ref{aff175}}
\and D.~Vergani\orcid{0000-0003-0898-2216}\inst{\ref{aff20}}
\and P.~Vielzeuf\orcid{0000-0003-2035-9339}\inst{\ref{aff14}}
\and E.~Vilenius\orcid{0000-0002-6184-7681}\inst{\ref{aff77}}
\and S.~Vinciguerra\orcid{0009-0005-4018-3184}\inst{\ref{aff56}}
\and M.~von~Wietersheim-Kramsta\orcid{0000-0003-4986-5091}\inst{\ref{aff84},\ref{aff149}}
\and N.~A.~Walton\orcid{0000-0003-3983-8778}\inst{\ref{aff170}}
\and L.~Wang\orcid{0000-0002-6736-9158}\inst{\ref{aff176},\ref{aff13}}
\and J.~R.~Weaver\orcid{0000-0003-1614-196X}\inst{\ref{aff177}}
\and A.~H.~Wright\orcid{0000-0001-7363-7932}\inst{\ref{aff146}}
\and D.~Scott\orcid{0000-0002-6878-9840}\inst{\ref{aff178}}}
										   
\institute{Universit\'e Lyon 1, CNRS/IN2P3, IP2I Lyon, Villeurbanne, France\label{aff1}
\and
Centre de Recherche Astrophysique de Lyon, UMR5574, CNRS, Universit\'e Claude Bernard Lyon 1, ENS de Lyon, 69230, Saint-Genis-Laval, France\label{aff2}
\and
ESAC/ESA, Camino Bajo del Castillo, s/n., Urb. Villafranca del Castillo, 28692 Villanueva de la Ca\~nada, Madrid, Spain\label{aff3}
\and
Leiden Observatory, Leiden University, Einsteinweg 55, 2333 CC Leiden, The Netherlands\label{aff4}
\and
Space Science Data Center, Italian Space Agency, via del Politecnico snc, 00133 Roma, Italy\label{aff5}
\and
Aerospace Logistics Technology Engineering Company, Corso Marche 79, 10146 Torino, Italy\label{aff6}
\and
Center for Space Science and Technology II, UMBC, NASA GSFC\label{aff7}
\and
NASA Goddard Space Flight Center, Greenbelt, MD 20771, USA\label{aff8}
\and
Center for Research and Exploration in Space Science and Technology, NASA/GSFC, Greenbelt, MD 20771, USA\label{aff9}
\and
INAF-Osservatorio Astronomico di Roma, Via Frascati 33, 00078 Monteporzio Catone, Italy\label{aff10}
\and
Caltech/IPAC, 1200 E. California Blvd., Pasadena, CA 91125, USA\label{aff11}
\and
INAF-Osservatorio Astronomico di Trieste, Via G. B. Tiepolo 11, 34143 Trieste, Italy\label{aff12}
\and
Kapteyn Astronomical Institute, University of Groningen, PO Box 800, 9700 AV Groningen, The Netherlands\label{aff13}
\and
Aix-Marseille Universit\'e, CNRS/IN2P3, CPPM, Marseille, France\label{aff14}
\and
Max-Planck-Institut f\"ur Astronomie, K\"onigstuhl 17, 69117 Heidelberg, Germany\label{aff15}
\and
California Institute of Technology, 1200 E California Blvd, Pasadena, CA 91125, USA\label{aff16}
\and
INAF-Osservatorio Astronomico di Padova, Via dell'Osservatorio 5, 35122 Padova, Italy\label{aff17}
\and
Infrared Processing and Analysis Center, California Institute of Technology, Pasadena, CA 91125, USA\label{aff18}
\and
INAF-Osservatorio Astronomico di Brera, Via Brera 28, 20122 Milano, Italy\label{aff19}
\and
INAF-Osservatorio di Astrofisica e Scienza dello Spazio di Bologna, Via Piero Gobetti 93/3, 40129 Bologna, Italy\label{aff20}
\and
Universit\'e Paris-Saclay, Universit\'e Paris Cit\'e, CEA, CNRS, AIM, 91191, Gif-sur-Yvette, France\label{aff21}
\and
IFPU, Institute for Fundamental Physics of the Universe, via Beirut 2, 34151 Trieste, Italy\label{aff22}
\and
INFN, Sezione di Trieste, Via Valerio 2, 34127 Trieste TS, Italy\label{aff23}
\and
SISSA, International School for Advanced Studies, Via Bonomea 265, 34136 Trieste TS, Italy\label{aff24}
\and
Dipartimento di Fisica e Astronomia, Universit\`a di Bologna, Via Gobetti 93/2, 40129 Bologna, Italy\label{aff25}
\and
INFN-Sezione di Bologna, Viale Berti Pichat 6/2, 40127 Bologna, Italy\label{aff26}
\and
Dipartimento di Fisica, Universit\`a di Genova, Via Dodecaneso 33, 16146, Genova, Italy\label{aff27}
\and
INFN-Sezione di Genova, Via Dodecaneso 33, 16146, Genova, Italy\label{aff28}
\and
Department of Physics "E. Pancini", University Federico II, Via Cinthia 6, 80126, Napoli, Italy\label{aff29}
\and
INAF-Osservatorio Astronomico di Capodimonte, Via Moiariello 16, 80131 Napoli, Italy\label{aff30}
\and
Instituto de Astrof\'isica e Ci\^encias do Espa\c{c}o, Universidade do Porto, CAUP, Rua das Estrelas, PT4150-762 Porto, Portugal\label{aff31}
\and
Faculdade de Ci\^encias da Universidade do Porto, Rua do Campo de Alegre, 4150-007 Porto, Portugal\label{aff32}
\and
European Southern Observatory, Karl-Schwarzschild-Str.~2, 85748 Garching, Germany\label{aff33}
\and
Dipartimento di Fisica, Universit\`a degli Studi di Torino, Via P. Giuria 1, 10125 Torino, Italy\label{aff34}
\and
INFN-Sezione di Torino, Via P. Giuria 1, 10125 Torino, Italy\label{aff35}
\and
INAF-Osservatorio Astrofisico di Torino, Via Osservatorio 20, 10025 Pino Torinese (TO), Italy\label{aff36}
\and
INAF-IASF Milano, Via Alfonso Corti 12, 20133 Milano, Italy\label{aff37}
\and
Centro de Investigaciones Energ\'eticas, Medioambientales y Tecnol\'ogicas (CIEMAT), Avenida Complutense 40, 28040 Madrid, Spain\label{aff38}
\and
Port d'Informaci\'{o} Cient\'{i}fica, Campus UAB, C. Albareda s/n, 08193 Bellaterra (Barcelona), Spain\label{aff39}
\and
Institute of Space Sciences (ICE, CSIC), Campus UAB, Carrer de Can Magrans, s/n, 08193 Barcelona, Spain\label{aff40}
\and
Institut d'Estudis Espacials de Catalunya (IEEC),  Edifici RDIT, Campus UPC, 08860 Castelldefels, Barcelona, Spain\label{aff41}
\and
INFN -- Sezione di Napoli, Via Cinthia 6, 80126, Napoli, Italy\label{aff42}
\and
Institute for Astronomy, University of Hawaii, 2680 Woodlawn Drive, Honolulu, HI 96822, USA\label{aff43}
\and
Dipartimento di Fisica e Astronomia "Augusto Righi" - Alma Mater Studiorum Universit\`a di Bologna, Viale Berti Pichat 6/2, 40127 Bologna, Italy\label{aff44}
\and
Instituto de Astrof\'{\i}sica de Canarias, E-38205 La Laguna, Tenerife, Spain\label{aff45}
\and
Institute for Astronomy, University of Edinburgh, Royal Observatory, Blackford Hill, Edinburgh EH9 3HJ, UK\label{aff46}
\and
Jodrell Bank Centre for Astrophysics, Department of Physics and Astronomy, University of Manchester, Oxford Road, Manchester M13 9PL, UK\label{aff47}
\and
European Space Agency/ESRIN, Largo Galileo Galilei 1, 00044 Frascati, Roma, Italy\label{aff48}
\and
Institut de Ci\`{e}ncies del Cosmos (ICCUB), Universitat de Barcelona (IEEC-UB), Mart\'{i} i Franqu\`{e}s 1, 08028 Barcelona, Spain\label{aff49}
\and
Instituci\'o Catalana de Recerca i Estudis Avan\c{c}ats (ICREA), Passeig de Llu\'{\i}s Companys 23, 08010 Barcelona, Spain\label{aff50}
\and
Institut de Ciencies de l'Espai (IEEC-CSIC), Campus UAB, Carrer de Can Magrans, s/n Cerdanyola del Vall\'es, 08193 Barcelona, Spain\label{aff51}
\and
Universit\'e Lyon 1, CNRS/IN2P3, IUF, IP2I Lyon, Villeurbanne, France \label{aff52}
\and
Departamento de F\'isica, Faculdade de Ci\^encias, Universidade de Lisboa, Edif\'icio C8, Campo Grande, PT1749-016 Lisboa, Portugal\label{aff53}
\and
Instituto de Astrof\'isica e Ci\^encias do Espa\c{c}o, Faculdade de Ci\^encias, Universidade de Lisboa, Campo Grande, 1749-016 Lisboa, Portugal\label{aff54}
\and
Department of Astronomy, University of Geneva, ch. d'Ecogia 16, 1290 Versoix, Switzerland\label{aff55}
\and
Aix-Marseille Universit\'e, CNRS, CNES, LAM, Marseille, France\label{aff56}
\and
Universit\'e Paris-Saclay, CNRS, Institut d'astrophysique spatiale, 91405, Orsay, France\label{aff57}
\and
INFN-Padova, Via Marzolo 8, 35131 Padova, Italy\label{aff58}
\and
INAF-Istituto di Astrofisica e Planetologia Spaziali, via del Fosso del Cavaliere, 100, 00100 Roma, Italy\label{aff59}
\and
School of Physics, HH Wills Physics Laboratory, University of Bristol, Tyndall Avenue, Bristol, BS8 1TL, UK\label{aff60}
\and
Department of Physics, University of Oxford, Keble Road, Oxford OX1 3RH, UK\label{aff61}
\and
University Observatory, LMU Faculty of Physics, Scheinerstr.~1, 81679 Munich, Germany\label{aff62}
\and
Max Planck Institute for Extraterrestrial Physics, Giessenbachstr. 1, 85748 Garching, Germany\label{aff63}
\and
Universit\"ats-Sternwarte M\"unchen, Fakult\"at f\"ur Physik, Ludwig-Maximilians-Universit\"at M\"unchen, Scheinerstr.~1, 81679 M\"unchen, Germany\label{aff64}
\and
Dipartimento di Fisica "Aldo Pontremoli", Universit\`a degli Studi di Milano, Via Celoria 16, 20133 Milano, Italy\label{aff65}
\and
INFN-Sezione di Milano, Via Celoria 16, 20133 Milano, Italy\label{aff66}
\and
Institute of Theoretical Astrophysics, University of Oslo, P.O. Box 1029 Blindern, 0315 Oslo, Norway\label{aff67}
\and
Jet Propulsion Laboratory, California Institute of Technology, 4800 Oak Grove Drive, Pasadena, CA, 91109, USA\label{aff68}
\and
Felix Hormuth Engineering, Goethestr. 17, 69181 Leimen, Germany\label{aff69}
\and
Technical University of Denmark, Elektrovej 327, 2800 Kgs. Lyngby, Denmark\label{aff70}
\and
Cosmic Dawn Center (DAWN), Denmark\label{aff71}
\and
Institut d'Astrophysique de Paris, UMR 7095, CNRS, and Sorbonne Universit\'e, 98 bis boulevard Arago, 75014 Paris, France\label{aff72}
\and
Universit\'e de Gen\`eve, D\'epartement de Physique Th\'eorique and Centre for Astroparticle Physics, 24 quai Ernest-Ansermet, CH-1211 Gen\`eve 4, Switzerland\label{aff73}
\and
Department of Physics, P.O. Box 64, University of Helsinki, 00014 Helsinki, Finland\label{aff74}
\and
Helsinki Institute of Physics, Gustaf H{\"a}llstr{\"o}min katu 2, University of Helsinki, 00014 Helsinki, Finland\label{aff75}
\and
Laboratoire d'etude de l'Univers et des phenomenes eXtremes, Observatoire de Paris, Universit\'e PSL, Sorbonne Universit\'e, CNRS, 92190 Meudon, France\label{aff76}
\and
Mullard Space Science Laboratory, University College London, Holmbury St Mary, Dorking, Surrey RH5 6NT, UK\label{aff77}
\and
SKAO, Jodrell Bank, Lower Withington, Macclesfield SK11 9FT, UK\label{aff78}
\and
Centre de Calcul de l'IN2P3/CNRS, 21 avenue Pierre de Coubertin 69627 Villeurbanne Cedex, France\label{aff79}
\and
University of Applied Sciences and Arts of Northwestern Switzerland, School of Computer Science, 5210 Windisch, Switzerland\label{aff80}
\and
Universit\"at Bonn, Argelander-Institut f\"ur Astronomie, Auf dem H\"ugel 71, 53121 Bonn, Germany\label{aff81}
\and
INFN-Sezione di Roma, Piazzale Aldo Moro, 2 - c/o Dipartimento di Fisica, Edificio G. Marconi, 00185 Roma, Italy\label{aff82}
\and
Dipartimento di Fisica e Astronomia "Augusto Righi" - Alma Mater Studiorum Universit\`a di Bologna, via Piero Gobetti 93/2, 40129 Bologna, Italy\label{aff83}
\and
Department of Physics, Institute for Computational Cosmology, Durham University, South Road, Durham, DH1 3LE, UK\label{aff84}
\and
Niels Bohr Institute, University of Copenhagen, Jagtvej 128, 2200 Copenhagen, Denmark\label{aff85}
\and
Universit\'e Paris Cit\'e, CNRS, Astroparticule et Cosmologie, 75013 Paris, France\label{aff86}
\and
CNRS-UCB International Research Laboratory, Centre Pierre Bin\'etruy, IRL2007, CPB-IN2P3, Berkeley, USA\label{aff87}
\and
Institute of Physics, Laboratory of Astrophysics, Ecole Polytechnique F\'ed\'erale de Lausanne (EPFL), Observatoire de Sauverny, 1290 Versoix, Switzerland\label{aff88}
\and
Dipartimento di Fisica - Sezione di Astronomia, Universit\`a di Trieste, Via Tiepolo 11, 34131 Trieste, Italy\label{aff89}
\and
Telespazio UK S.L. for European Space Agency (ESA), Camino bajo del Castillo, s/n, Urbanizacion Villafranca del Castillo, Villanueva de la Ca\~nada, 28692 Madrid, Spain\label{aff90}
\and
Institut de F\'{i}sica d'Altes Energies (IFAE), The Barcelona Institute of Science and Technology, Campus UAB, 08193 Bellaterra (Barcelona), Spain\label{aff91}
\and
School of Mathematics and Physics, University of Surrey, Guildford, Surrey, GU2 7XH, UK\label{aff92}
\and
European Space Agency/ESTEC, Keplerlaan 1, 2201 AZ Noordwijk, The Netherlands\label{aff93}
\and
Waterloo Centre for Astrophysics, University of Waterloo, Waterloo, Ontario N2L 3G1, Canada\label{aff94}
\and
Department of Physics and Astronomy, University of Waterloo, Waterloo, Ontario N2L 3G1, Canada\label{aff95}
\and
Perimeter Institute for Theoretical Physics, Waterloo, Ontario N2L 2Y5, Canada\label{aff96}
\and
Institute of Space Science, Str. Atomistilor, nr. 409 M\u{a}gurele, Ilfov, 077125, Romania\label{aff97}
\and
Consejo Superior de Investigaciones Cientificas, Calle Serrano 117, 28006 Madrid, Spain\label{aff98}
\and
Universidad de La Laguna, Dpto. Astrof\'\i sica, E-38206 La Laguna, Tenerife, Spain\label{aff99}
\and
Institute for Particle Physics and Astrophysics, Dept. of Physics, ETH Zurich, Wolfgang-Pauli-Strasse 27, 8093 Zurich, Switzerland\label{aff100}
\and
Dipartimento di Fisica e Astronomia "G. Galilei", Universit\`a di Padova, Via Marzolo 8, 35131 Padova, Italy\label{aff101}
\and
Instituto de F\'isica Te\'orica UAM-CSIC, Campus de Cantoblanco, 28049 Madrid, Spain\label{aff102}
\and
Institut de Recherche en Astrophysique et Plan\'etologie (IRAP), Universit\'e de Toulouse, CNRS, UPS, CNES, 14 Av. Edouard Belin, 31400 Toulouse, France\label{aff103}
\and
Universit\'e St Joseph; Faculty of Sciences, Beirut, Lebanon\label{aff104}
\and
Departamento de F\'isica, FCFM, Universidad de Chile, Blanco Encalada 2008, Santiago, Chile\label{aff105}
\and
Department of Physics and Helsinki Institute of Physics, Gustaf H\"allstr\"omin katu 2, University of Helsinki, 00014 Helsinki, Finland\label{aff106}
\and
Instituto de Astrof\'isica e Ci\^encias do Espa\c{c}o, Faculdade de Ci\^encias, Universidade de Lisboa, Tapada da Ajuda, 1349-018 Lisboa, Portugal\label{aff107}
\and
Cosmic Dawn Center (DAWN)\label{aff108}
\and
Universidad Polit\'ecnica de Cartagena, Departamento de Electr\'onica y Tecnolog\'ia de Computadoras,  Plaza del Hospital 1, 30202 Cartagena, Spain\label{aff109}
\and
European University of Technology EUt+, European Union\label{aff110}
\and
Carnegie Observatories, Pasadena, CA 91101, USA\label{aff111}
\and
Dipartimento di Fisica e Scienze della Terra, Universit\`a degli Studi di Ferrara, Via Giuseppe Saragat 1, 44122 Ferrara, Italy\label{aff112}
\and
Istituto Nazionale di Fisica Nucleare, Sezione di Ferrara, Via Giuseppe Saragat 1, 44122 Ferrara, Italy\label{aff113}
\and
ICSC - Centro Nazionale di Ricerca in High Performance Computing, Big Data e Quantum Computing, Via Magnanelli 2, Bologna, Italy\label{aff114}
\and
INAF, Istituto di Radioastronomia, Via Piero Gobetti 101, 40129 Bologna, Italy\label{aff115}
\and
INFN-Bologna, Via Irnerio 46, 40126 Bologna, Italy\label{aff116}
\and
Astronomical Observatory of the Autonomous Region of the Aosta Valley (OAVdA), Loc. Lignan 39, I-11020, Nus (Aosta Valley), Italy\label{aff117}
\and
Department of Mathematical Physics, Institute of Physics, University of S\~ao Paulo, R. do Mat\~ao 1371, 05508-090, S\~ao Paulo, SP, Brazil\label{aff118}
\and
CEA Saclay, DFR/IRFU, Service d'Astrophysique, Bat. 709, 91191 Gif-sur-Yvette, France\label{aff119}
\and
Code 665, NASA/GSFC, 8800 Greenbelt Road, Greenbelt, MD 20771, USA\label{aff120}
\and
SSAI, Lanham, MD 20706, USA\label{aff121}
\and
Department of Astronomy, University of Maryland, College Park, MD 20742, USA\label{aff122}
\and
Oskar Klein Centre for Cosmoparticle Physics, Department of Physics, Stockholm University, Stockholm, SE-106 91, Sweden\label{aff123}
\and
Astrophysics Group, Blackett Laboratory, Imperial College London, London SW7 2AZ, UK\label{aff124}
\and
Univ. Grenoble Alpes, CNRS, Grenoble INP, LPSC-IN2P3, 53, Avenue des Martyrs, 38000 Grenoble, France\label{aff125}
\and
Dipartimento di Fisica, Sapienza Universit\`a di Roma, Piazzale Aldo Moro 2, 00185 Roma, Italy\label{aff126}
\and
Department of Mathematics and Physics E. De Giorgi, University of Salento, Via per Arnesano, CP-I93, 73100, Lecce, Italy\label{aff127}
\and
INFN, Sezione di Lecce, Via per Arnesano, CP-193, 73100, Lecce, Italy\label{aff128}
\and
INAF-Sezione di Lecce, c/o Dipartimento Matematica e Fisica, Via per Arnesano, 73100, Lecce, Italy\label{aff129}
\and
Institut d'Astrophysique de Paris, 98bis Boulevard Arago, 75014, Paris, France\label{aff130}
\and
ICL, Junia, Universit\'e Catholique de Lille, LITL, 59000 Lille, France\label{aff131}
\and
CERCA/ISO, Department of Physics, Case Western Reserve University, 10900 Euclid Avenue, Cleveland, OH 44106, USA\label{aff132}
\and
Laboratoire Univers et Th\'eorie, Observatoire de Paris, Universit\'e PSL, Universit\'e Paris Cit\'e, CNRS, 92190 Meudon, France\label{aff133}
\and
Departamento de F{\'\i}sica Fundamental. Universidad de Salamanca. Plaza de la Merced s/n. 37008 Salamanca, Spain\label{aff134}
\and
Universiteit Gent, Department of Physics and Astronomy, Proeftuinstraat 86 N3, 9000 Ghent, Belgium
\label{aff135}
\and
Universit\'e de Strasbourg, CNRS, Observatoire astronomique de Strasbourg, UMR 7550, 67000 Strasbourg, France\label{aff136}
\and
Center for Data-Driven Discovery, Kavli IPMU (WPI), UTIAS, The University of Tokyo, Kashiwa, Chiba 277-8583, Japan\label{aff137}
\and
Institute of Cosmology and Gravitation, University of Portsmouth, Portsmouth PO1 3FX, UK\label{aff138}
\and
Department of Physics \& Astronomy, University of California Irvine, Irvine CA 92697, USA\label{aff139}
\and
Universit\"at Innsbruck, Institut f\"ur Astro- und Teilchenphysik, Technikerstr. 25/8, 6020 Innsbruck, Austria\label{aff140}
\and
Departamento F\'isica Aplicada, Universidad Polit\'ecnica de Cartagena, Campus Muralla del Mar, 30202 Cartagena, Murcia, Spain\label{aff141}
\and
Instituto de F\'isica de Cantabria, Edificio Juan Jord\'a, Avenida de los Castros, 39005 Santander, Spain\label{aff142}
\and
Department of Astronomy, University of Florida, Bryant Space Science Center, Gainesville, FL 32611, USA\label{aff143}
\and
Department of Computer Science, Aalto University, PO Box 15400, Espoo, FI-00 076, Finland\label{aff144}
\and
School of Physics and Astronomy, University of Nottingham, University Park, Nottingham NG7 2RD, UK\label{aff145}
\and
Ruhr University Bochum, Faculty of Physics and Astronomy, Astronomical Institute (AIRUB), German Centre for Cosmological Lensing (GCCL), 44780 Bochum, Germany\label{aff146}
\and
DARK, Niels Bohr Institute, University of Copenhagen, Jagtvej 155, 2200 Copenhagen, Denmark\label{aff147}
\and
INAF-Osservatorio Astrofisico di Arcetri, Largo E. Fermi 5, 50125, Firenze, Italy\label{aff148}
\and
Department of Physics, Centre for Extragalactic Astronomy, Durham University, South Road, Durham, DH1 3LE, UK\label{aff149}
\and
Astrophysics Research Centre, University of KwaZulu-Natal, Westville Campus, Durban 4041, South Africa\label{aff150}
\and
School of Mathematics, Statistics \& Computer Science, University of KwaZulu-Natal, Westville Campus, Durban 4041, South Africa\label{aff151}
\and
Department of Physics and Astronomy, Vesilinnantie 5, University of Turku, 20014 Turku, Finland\label{aff152}
\and
Finnish Centre for Astronomy with ESO (FINCA), Quantum, Vesilinnantie 5, University of Turku, 20014 Turku, Finland\label{aff153}
\and
Serco for European Space Agency (ESA), Camino bajo del Castillo, s/n, Urbanizacion Villafranca del Castillo, Villanueva de la Ca\~nada, 28692 Madrid, Spain\label{aff154}
\and
ARC Centre of Excellence for Dark Matter Particle Physics, Melbourne, Australia\label{aff155}
\and
Centre for Astrophysics \& Supercomputing, Swinburne University of Technology,  Hawthorn, Victoria 3122, Australia\label{aff156}
\and
Department of Physics and Astronomy, University of the Western Cape, Bellville, Cape Town, 7535, South Africa\label{aff157}
\and
MTA-CSFK Lend\"ulet Large-Scale Structure Research Group, Konkoly-Thege Mikl\'os \'ut 15-17, H-1121 Budapest, Hungary\label{aff158}
\and
Konkoly Observatory, HUN-REN CSFK, MTA Centre of Excellence, Budapest, Konkoly Thege Mikl\'os {\'u}t 15-17. H-1121, Hungary\label{aff159}
\and
Institute for Theoretical Particle Physics and Cosmology (TTK), RWTH Aachen University, 52056 Aachen, Germany\label{aff160}
\and
IRFU, CEA, Universit\'e Paris-Saclay 91191 Gif-sur-Yvette Cedex, France\label{aff161}
\and
Centro de Astrof\'{\i}sica da Universidade do Porto, Rua das Estrelas, 4150-762 Porto, Portugal\label{aff162}
\and
HE Space for European Space Agency (ESA), Camino bajo del Castillo, s/n, Urbanizacion Villafranca del Castillo, Villanueva de la Ca\~nada, 28692 Madrid, Spain\label{aff163}
\and
Department of Astrophysics, University of Zurich, Winterthurerstrasse 190, 8057 Zurich, Switzerland\label{aff164}
\and
Theoretical astrophysics, Department of Physics and Astronomy, Uppsala University, Box 524, 751 20 Uppsala, Sweden\label{aff165}
\and
Mathematical Institute, University of Leiden, Einsteinweg 55, 2333 CA Leiden, The Netherlands\label{aff166}
\and
School of Physical Sciences, The Open University, Milton Keynes, MK7 6AA, UK\label{aff167}
\and
ASTRON, the Netherlands Institute for Radio Astronomy, Postbus 2, 7990 AA, Dwingeloo, The Netherlands\label{aff168}
\and
Center for Advanced Interdisciplinary Research, Ss. Cyril and Methodius University in Skopje, Macedonia\label{aff169}
\and
Institute of Astronomy, University of Cambridge, Madingley Road, Cambridge CB3 0HA, UK\label{aff170}
\and
Center for Astrophysics and Cosmology, University of Nova Gorica, Nova Gorica, Slovenia\label{aff171}
\and
Department of Astrophysical Sciences, Peyton Hall, Princeton University, Princeton, NJ 08544, USA\label{aff172}
\and
Fakult\"at f\"ur Physik, Universit\"at Bielefeld, Postfach 100131, 33501 Bielefeld, Germany\label{aff173}
\and
School of Mathematics, Statistics and Physics, Newcastle University, Herschel Building, Newcastle-upon-Tyne, NE1 7RU, UK\label{aff174}
\and
Space physics and astronomy research unit, University of Oulu, Pentti Kaiteran katu 1, FI-90014 Oulu, Finland\label{aff175}
\and
SRON Netherlands Institute for Space Research, Landleven 12, 9747 AD, Groningen, The Netherlands\label{aff176}
\and
MIT Kavli Institute for Astrophysics and Space Research, Massachusetts Institute of Technology, Cambridge, MA 02139, USA\label{aff177}
\and
Department of Physics and Astronomy, University of British Columbia, Vancouver, BC V6T 1Z1, Canada\label{aff178}}    
\date{}

\titlerunning{Persistence in NISP detectors}
\authorrunning{Euclid Collaboration: B. Kubik et al.}

\abstract{
Detector persistence is a significant source of systematic error in infrared observations and affects the scientific performance of the \Euclid Near Infrared Spectrometer and Photometer (NISP). We characterise persistence using complementary ground calibration and in-flight data, develop empirical models of its temporal evolution, and assess its impact on science observations and the \Euclid data-processing pipeline. Persistence is highly non-uniform across the focal plane, with variations of up to two orders of magnitude between and within detectors. Its amplitude is typically below 1\% of the stimulus signal and depends nonlinearly on the preceding fluence. The decay rate is faster for brighter sources. During mission operations, the persistence signal has progressively decreased by roughly 10\%, while the decay rate has remained approximately stable. Persistence following over-saturating signals does not follow the empirical relations derived for under-saturating fluences and plausibly reduces persistence response over longer timescales. The impact of charge trapping and release on flat-field observations and on the background in science exposures is also discussed. Persistence measured in science observations is observed to differ systematically from that derived from dark calibration data. In addition, persistence of energetic particle hits is observed to be different from that of astronomical sources of comparable signal. Finally, we discuss the impact of persistence on \Euclid NISP observations and describe the masking strategy adopted for the Q1 abd DR1 releases, focusing on identifying which pixels are affected by persistence and for how long they should be excluded from science observations.
}


\maketitle

\section{Introduction}
\label{sec:intro}

Persistence is a well-known systematic effect affecting Source Follower HgCdTe infrared detectors (i.e. an array of $N \times N$ photosensitive pixels) used in astronomy, whereby residual signal from prior illumination contaminates subsequent exposures. This signal, typically nonlinear and often significantly larger than the nominal dark current, can persist over timescales of several hours. The phenomenon is generally attributed to defects and impurities in the semiconductor material, which introduce localized energy states within the bandgap that act as carrier traps. It is not clear though how many trap species contribute under which conditions to the persistence effect. We assume for the Euclid H2RGs a wide range of release timescales. A widely used phenomenological framework attributes the effect to charge trapping and release within the pixel, modelled as a photodiode with a variable depletion region \citep{Smith_2008a,Smith_2008b}. Three-dimensional trap density maps obtained for James Webb Space Telescope H2RG\footnote{The H2RG detector belongs to the HAWAII family of HgCdTe infrared detector arrays, with HAWAII standing for HgCdTe Astronomical Wide Area Infrared Imager. In the HxRG nomenclature, H denotes HAWAII,  x=1,\,2,\,4 corresponds to 1024\,$\times$\,1024, 2048\,$\times$\,2048 or 4096\,$\times$\,4096 pixels, respectively, while R and G denote reference pixels and guide-window capability.} detectors show that traps are not uniformly distributed throughout the depletion region \citep{Anderson_2013}. Subsequent studies have characterised the persistence amplitude and decay as functions of flux, fluence, or soak time, i.e. the time over which a given charge remains in the pixel.

From an observational standpoint, persistence is typically described phenomenologically as a long-lived, illumination-dependent signal affecting subsequent exposures. Both exponential \citep{Serra_2015,Tulloch_2019,Mosby_2020} and power-law \citep{Long_2012,Long_2015} decay models have been proposed, the latter suggesting a broad distribution of trapping timescales. While such empirical descriptions are sufficient for calibration purposes, they do not provide direct insight into the underlying physical mechanisms, which remain an open question.

Persistence introduces memory effects in the data, imprinting previous exposures onto subsequent images and thereby complicating both calibration and data reduction. Under nominal observing conditions, any perturbation of the detector -- such as illumination, reverse bias variations, or saturation -- can induce persistence. Its long characteristic decay time makes calibration particularly challenging, since measurements may require hours or even days to return to equilibrium. Unlike dark current, for which standardised calibration procedures exist across missions, persistence lacks a unified framework; calibration strategies must therefore be tailored to the specific scientific objectives and operational constraints of each instrument. As a result, persistence affects multiple aspects of instrument operation, including calibration planning, observing strategies, and data processing pipelines.

The aim of this paper is to summarise the efforts undertaken by the authors within the \Euclid mission to characterise and calibrate persistence, and to describe the data reduction pipeline developed to mitigate its impact on the science data. We introduce a phenomenological model for long-lived persistence signals, calibrated using both ground-based characterisation data and in-flight observations, and we discuss its domain of applicability. We also briefly examine the observed persistence behaviour under various observing conditions in the context of existing models. A detailed physical interpretation, however, is beyond the scope of this work, since limited information is available on detector architecture, material composition, and fabrication processes, all of which may significantly influence persistence properties.

The paper is organised as follows. In Sect.~\ref{sec:readout} we describe the \Euclid observing sequence and readout modes, which set the framework for persistence calibration and masking. In Sect.~\ref{sec:data} we present the data sets used for characterisation, including ground-based and in-flight calibration data, as well as nominal observations. In Sect.~\ref{sec:calibration} we introduce the phenomenological model of persistence decay. Owing to differences in data acquisition, we adopt two complementary parameterisations based on ground and in-flight data (Sect.~\ref{sec:parameterisation}). We then describe the spatial distribution of persistence across the focal plane (Sect.~\ref{sec:regions}), and its dependence on prior signal level (Sect.~\ref{sec:signal_dependence}) and time (Sect.~\ref{sec:time_evolution}). The effects of strong saturation (Sect.~\ref{sec:saturation}) and the impact of persistence on flat-field stability (Sect.~\ref{sec:impact_on_flats}) are also discussed. In Sect.~\ref{sec:pers_sience} we compare persistence measured in calibration data with that observed in nominal conditions, including contributions from astrophysical sources and energetic particle hits (EPHs), and we assess its impact on background estimation. The limitations of the model are summarised in Sect.~\ref{sec:limitations}. Finally, in Sect.~\ref{sec:application}, we describe its implementation in the data reduction pipelines, including the processing adopted for the Q1 and DR1 releases (Sect.~\ref{sec:pipelines}) and subsequent optimisations (Sect.~\ref{sec:optimisation}), before concluding in Sect.~\ref{sec:conclusions}.

\section{NISP observing sequence and detector readout}
\label{sec:readout}

During \Euclid's nominal survey, NISP continuously scans the sky following a predefined Reference Observation Sequence \citep[ROS,][]{Scaramella-EP1}. Each ROS covers an area of $0.53$\,\sqdeg through four nearby pointings of a given field and consists of four dithers, each lasting approximately 15 minutes. Within each dither, a spectroscopic exposure of about 9 minutes is obtained using two grisms covering the 1206--1892 nm wavelength range. The grisms are mounted on a grism wheel and have dispersion directions differing by \ang{180}. In addition, the grism wheel is tilted by \ang{4} during the observation sequence, resulting in four distinct dispersion orientations for each observation \citep{Schirmer-EP18,EuclidSkyNISP,EP-Gillard}. Each spectroscopic exposure is followed by three photometric exposures of approximately 2 minutes each, obtained sequentially through the \JE (1168--1567 nm), \HE (1522--2021 nm), and \YE (950--1212 nm) filters, which are mounted on a filter wheel. During nominal survey observations, the NISP instrument operates in two predefined readout modes, optimised for spectroscopic and photometric measurements, respectively \citep{EuclidSkyNISP}. In both modes, the detectors measure the signal using the multi-accumulation (MACC) sampling. An exposure begins with a reset frame and then the signal is sampled up-the-ramp with $n_{\rm g}$ groups of $n_{\rm f}$ averaged frames. The groups are separated by $n_{\rm d}$ dropped frames \citep[see figure 2 in][]{EU-Kubik}. The readout time of a single frame is $t_{\rm fr} = 1.45408\,\mathrm{s}$. The spectroscopic mode uses MACC($n_{\rm g}$=15, $n_{\rm f}$=16, $n_{\rm d}$=11), resulting in a total exposure time of \SI{574.4}{s} (\SI{549.6}{s} integration time). The photometric mode employs MACC(4,16,4) with a total exposure time of 112.0\,s and integration time of \SI{87.2}{s} \citep[for details, see table\,1 in][]{EU-Kubik}. The signal accumulated in a given pixel during an exposure is estimated from the MACC samples using an algorithm implemented in the on-board electronics \citep{Kubik2016}. This algorithm simultaneously estimates the signal and evaluates the goodness of fit, quantified by the quality factor (QF), which enables the identification of anomalies occurring during the integration, such as EPH hits. These observing schemes, together with the signal estimation algorithm, determine the temporal sampling of the persistence signal and introduce an effective temporal averaging through the signal-estimation process.

\section{Data and terminology}
\label{sec:data}

In this paper we use a combination of ground calibration measurements and in-flight calibration data sets to characterise the persistence properties of the detectors over a wide range of illumination levels and wavelengths, and to monitor the persistence behaviour over time.  All data sets used in this work are summarised in Table~\ref{tab:datasets}, where the most relevant observational parameters are listed together with references to the sections in which each data set is used in the analysis. 
%
\begin{table*}
   \caption{Scientific objectives of the data sets used in this work. The table lists the most relevant observational parameters and references to the sections where each data set is used in the analysis. In the table, "all" means a measurement at each of the five calibration LED wavelengths, 939\,nm, 1157\,nm, 1467\,nm, 1735\,nm, and 1873\,nm. }
    \label{tab:datasets}
    \begin{center}       
    \begin{tabular}{lllllllll}
    \toprule
    Data set     & Date & Periodicity & LED & Fluence  & Flux & Primary objective & Sections \\
                &      &             &     & [ke$^-$] & [e$^-$s$^{-1}$] & & \\
    \midrule
    TV1         & 2019 & -           & 1467\,nm & 5--400 & 60--740 & persistence modelling & \ref{sec:signal_dependence}, \ref{sec:saturation} \\ 
    PV-010      & 2023 & -           & all      & 8--64 & 90--730 & flat-field stabilisation & \ref{sec:impact_on_flats} \\
    PV-016      & 2023 & -           & 1873\,nm & 20--100 & 230--1150 & persistence calibration  & \ref{sec:signal_dependence}, \ref{sec:time_evolution}\\
    F-007       & 2024 & monthly     & all      & 8 & 90 & persistence monitoring & \ref{sec:time_evolution} \\ 
    F-018       & 2025 & monthly     & 1873\,nm & 60 & 690 & persistence monitoring & \ref{sec:time_evolution} \\
    SkyPatch 38 & 2024 & - & - & - & - & persistence in science data & \ref{sec:pers_sience} \\
    \bottomrule
    \end{tabular}
    \end{center}
\end{table*} 

\subsection{Terminology}

Throughout this work, the exposure responsible for generating persistence in subsequent exposures is referred to as the `source exposure'. The `fluence' denotes the total integrated charge accumulated during an exposure (in electrons or ADU), whereas the `flux' is defined as the fluence divided by the integration time. The quantities `stimulus fluence' and `stimulus flux' refer specifically to the fluence and flux of the source exposure, regardless of whether that exposure is a flat-field or a nominal science observation. 

The exposure in which the persistence signal is evaluated is referred to as the `target exposure'. In practical applications, the target exposure is the frame for which the persistence contribution is estimated and subsequently masked.

In this work, the persistence charge $P$ accumulated during a target exposure is defined as the integral of the persistence current $I_{\rm P}$ over the exposure integration time. All quoted values of $P$ are computed assuming the photometric integration time. For spectroscopic exposures, the corresponding persistence charge must be rescaled to account for the different integration time. 

In this analysis, we consider persistence on a per-pixel basis in electrons or in ADU, as this is the relevant criterion for determining whether individual pixels should be masked due to persistence effects and, consequently, for assessing their impact on the total number of operable pixels as discussed in \citet{DR1-TP003}.

\subsection{Ground calibration data -- TV1}
\label{sec:data_tv1}

Ground calibration data from the Thermal Vacuum (TV1)  campaign, presented in \citet{EU-Kubik}, were used to characterise the persistence response over a wide range of fluences. Fluence levels range from 5 to 100\,ke$^{-}$ below saturation, complemented by two additional fluences of 300 and 400\,ke$^{-}$, well above the pixel full-well capacity of approximately 130\,ke$^{-}$. 

For fluence levels below saturation, the persistence current was measured during a dark exposure sampled up-the-ramp with 276 frames, UTR(276), following a flat-field exposure sampled with 76 frames, UTR(76). For the two oversaturated fluence levels, the persistence current was likewise measured during a dark exposure in UTR(276), but following a longer flat-field exposure in UTR(394). At each fluence level, the measurement was repeated in 15 consecutive realisations.

\subsection{Flight calibration data}
\label{sec:data_flight}

In-flight calibration data sets complement the ground measurements by probing persistence behaviour under operational conditions. In this work we use calibration data acquired during the performance verification phase \citep[PV,][]{DR1-TP026} and during the nominal operation flight phase \citep[F,][]{EuclidSkyOverview}. 

\paragraph{PV-010}

The PV-010 observations were performed in September 2023. For each of the five LEDs, labelled A (939\,nm), B (1157\,nm), C (1467\,nm), D (1735\,nm), and E (1873\,nm), \citep{EuclidSkyNISPCU}, a sequence of five fluence levels in flat-field illumination time was applied, with fluence progressively increasing from 8 to 64\,ke$^-$ (8, 18, 32, 50, 64\,ke$^-$). For each fluence, the flux level was maintained constant for approximately one hour, during which 30 exposures were acquired in photometric readout mode. In this work we use these flat-field exposures at each LED wavelength and fluence level to evaluate the impact of charge trapping on the flat-field stabilisation.

\paragraph{PV-016}

The PV-016 data were obtained in September 2023. This data set consists of persistence measurements acquired for fluence levels of 20, 40, 60, 80, and 100\,ke$^-$, applied in increasing order. Illumination was provided using LED E. After each fluence level, persistence decay was sampled using a sequence of dark exposures. The first five exposures were acquired in photometric readout mode MACC(4,16,4) with an exposure time of $t_{\rm exp} = 112.0\,\mathrm{s}$. The subsequent ten exposures were obtained in spectroscopic readout mode MACC(15,16,11) with $t_{\rm exp} = 574.4\,\mathrm{s}$. This sequence samples the decay over a total duration of approximately 1\,h\,40\,min for the first four fluence levels, extending up to 5\,h for the highest fluence level.

\paragraph{F-007}

The F-007 observations have been performed monthly since March 2024. Although these observations were not specifically designed to monitor persistence, this data set provides regular measurements of persistence amplitude during survey operations. For each LED (A--E), a single fluence level of 8\,ke$^-$ is applied, and probed in a sequence of 30 exposures acquired in photometric readout mode. The full flat-field sequence is followed by approximately 4.5\,h of dark exposures in spectroscopic readout mode. In this work, only the first dark exposure following each flat-field sequence is used to track the evolution of persistence amplitude during the survey.

\paragraph{F-018}

The F-018 data set comprises repeated persistence measurements collected monthly since March 2025. Each sequence begins with an illumination at a fluence level of 60\,ke$^-$ using LED E, followed by a dark phase lasting approximately 50 minutes. The decay is sampled using dark exposures obtained in photometric readout mode for the first five exposures, followed by four spectroscopic exposures. Each month, the full sequence -- consisting of flat-field illumination and dark exposures -- is performed twice consecutively.

\subsection{Flight nominal data -- SkyPatch38}
\label{sec:data_skypatch38}

SkyPatch38 consists of 43 observations obtained in the Euclid Deep Field North \citep[EDF-N,][]{EuclidSkyOverview} between 2024-06-12 and 2024-06-14. These exposures were affected by enhanced EPH showers associated with elevated solar activity. This data set provides a test case to characterise persistence induced by different sources, including stars, galaxies, and EPHs and enables a direct comparison of persistence measured directly on science data with the persistence model based on calibration observations.

\section{Persistence calibration}
\label{sec:calibration}

\subsection{Data processing and parameterisation}
\label{sec:parameterisation}

\paragraph{Ground parameterisation} 
During the ground calibration campaign, persistence was modelled using full UTR sequences consisting of 296 non-destructive readouts that were acquired. The ramps were sampled over approximately 7 minutes with a temporal resolution of $t_{\rm fr} = 1.45408\,\mathrm{s}$, enabling a detailed characterisation of the persistence evolution as a function of time. The temporal evolution of the persistent current, $I_{\rm P}$ was parameterised using an empirical model that provided the best fit to the experimental data as
\begin{equation}
    I_{\rm P}(S,t) = \alpha(S)\left(\frac{\tau}{t - t_0 + \tau}\right)^{\beta(S)}\,,
\label{eq:I}
\end{equation}
where $\alpha$ is the initial persistence current amplitude after a signal $S$, $t_0$ corresponds to the moment when the stimulus ends and the pixel is reset to its baseline level, and $\beta$ describes the temporal decay of the persistence current. The parameter $\tau$ is introduced to prevent a divergence at $t=t_0$ and effectively accounts for the $\sim 10\,\mu\rm{sec}$ delay between the end of the illumination and reset of the pixel. The persistence parameters $\alpha$ and $\beta$ were calibrated by fitting the time integral of Eq.~(\ref{eq:I}) to the accumulating persistence signal over the exposure duration at each fluence level $S$ \citep{Kubik24}.

\paragraph{Flight parameterisation} 
In contrast to the ground data, persistence measurements in flight are obtained from signals integrated over individual exposures, as described in \citet{Kubik2016} and \citet{EU-Kubik}, rather than from all the accessible frames in UTR sampling. Typically, the first five dark exposures are acquired using photometric readout mode with $t_{\rm exp} = 112.0\,\mathrm{s}$, while the remaining  exposures are taken using spectroscopic readout mode with $t_{\rm exp} = 574.4\,\mathrm{s}$.  This acquisition scheme makes the measurement less sensitive to fast persistence components, while providing a more precise characterisation of the long-term decay. Consequently, the model derived from in-flight calibration adopts a simplified functional form compared to the ground-based model,
\begin{equation} 
    I_{\rm P}(S,t)  = \tilde{\alpha}(S)\,(t-\tilde{t}_0)^{-\tilde{\beta}(S)}\,,
\label{eq:Ipip}
\end{equation}
where $\tilde{\alpha}$ and $\tilde{\beta}$ denote the initial persistence current amplitude and the decay index, respectively, $\tilde{t}_0$ is the  midpoint of the source exposure. The tildes emphasise that these coefficients are not directly comparable to those obtained from ground-based measurements, due to differences in the data acquisition scheme and the adopted parameterisation. Ground measurements provide higher precision and temporal resolution, but primarily probe only the fast component (around 7 minutes) of the persistence decay, whereas the in-flight data allow the evolution of the persistence signal to be tracked over longer time scales (around 1 hour) but with lower temporal resolution.

\subsection{Intra-detector regions}
\label{sec:regions}

\begin{figure*}
    \centering
    \includegraphics[width = 0.4\linewidth]{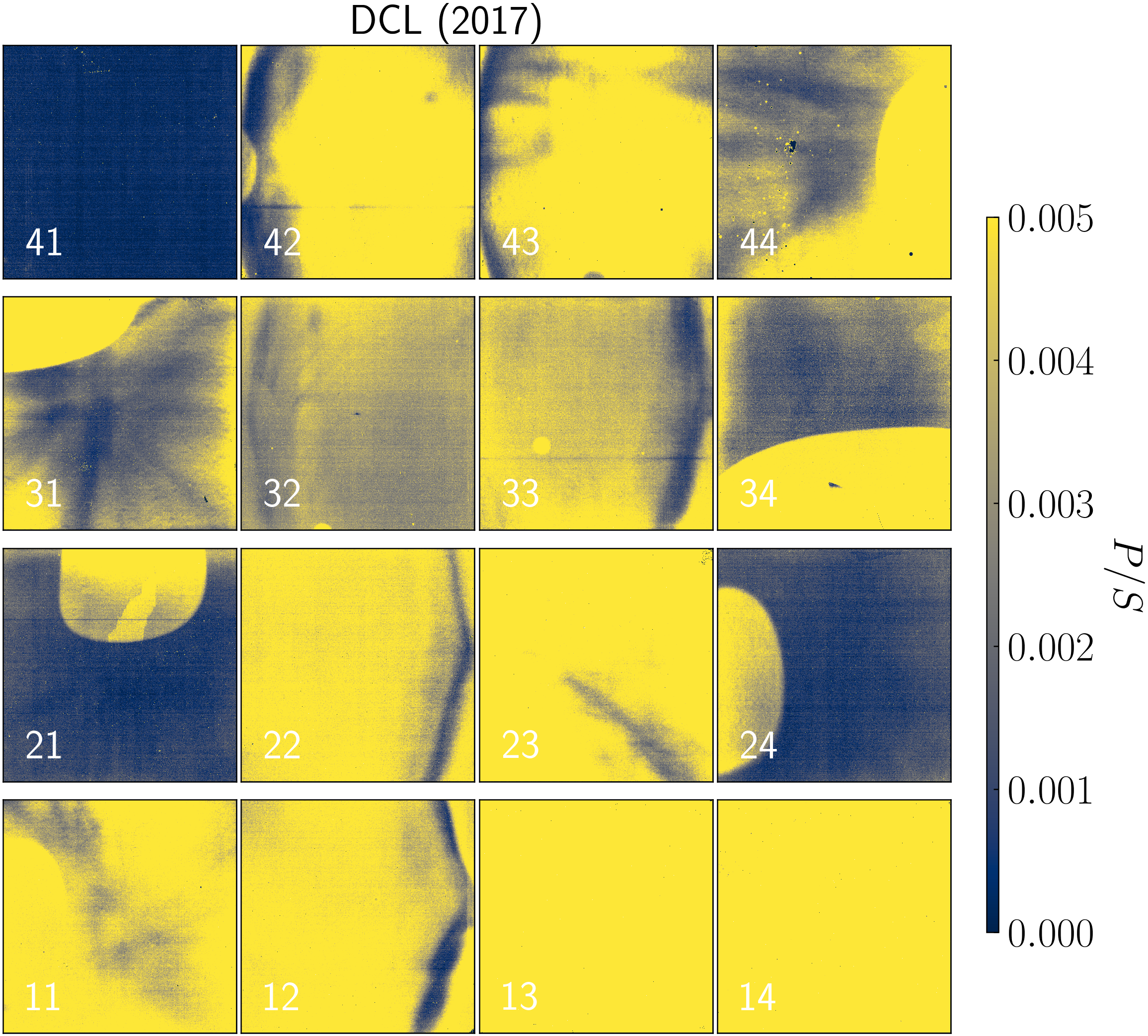}\qquad
    \includegraphics[width = 0.4\linewidth]{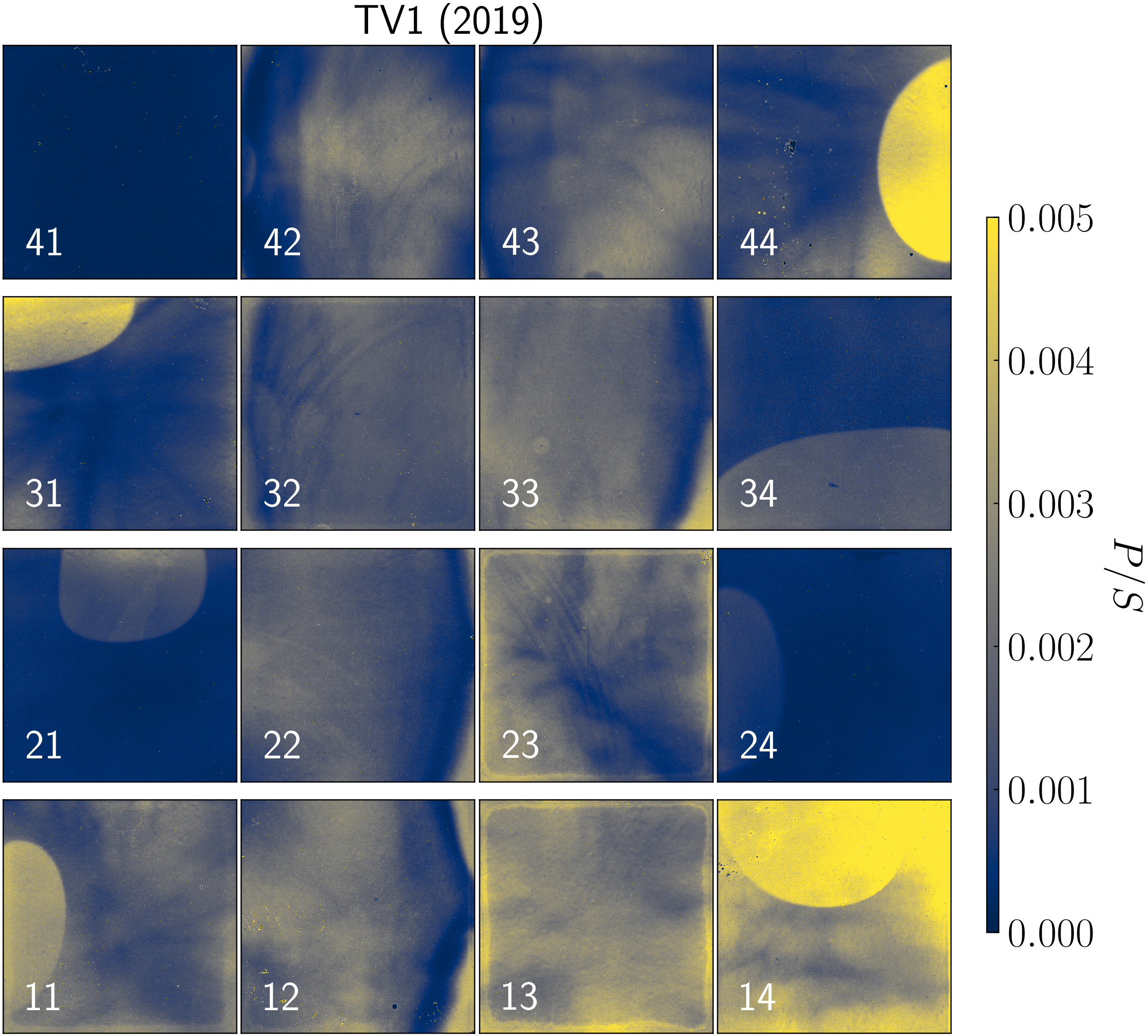} \vspace{1cm}\\
    \includegraphics[width = 0.4\linewidth]{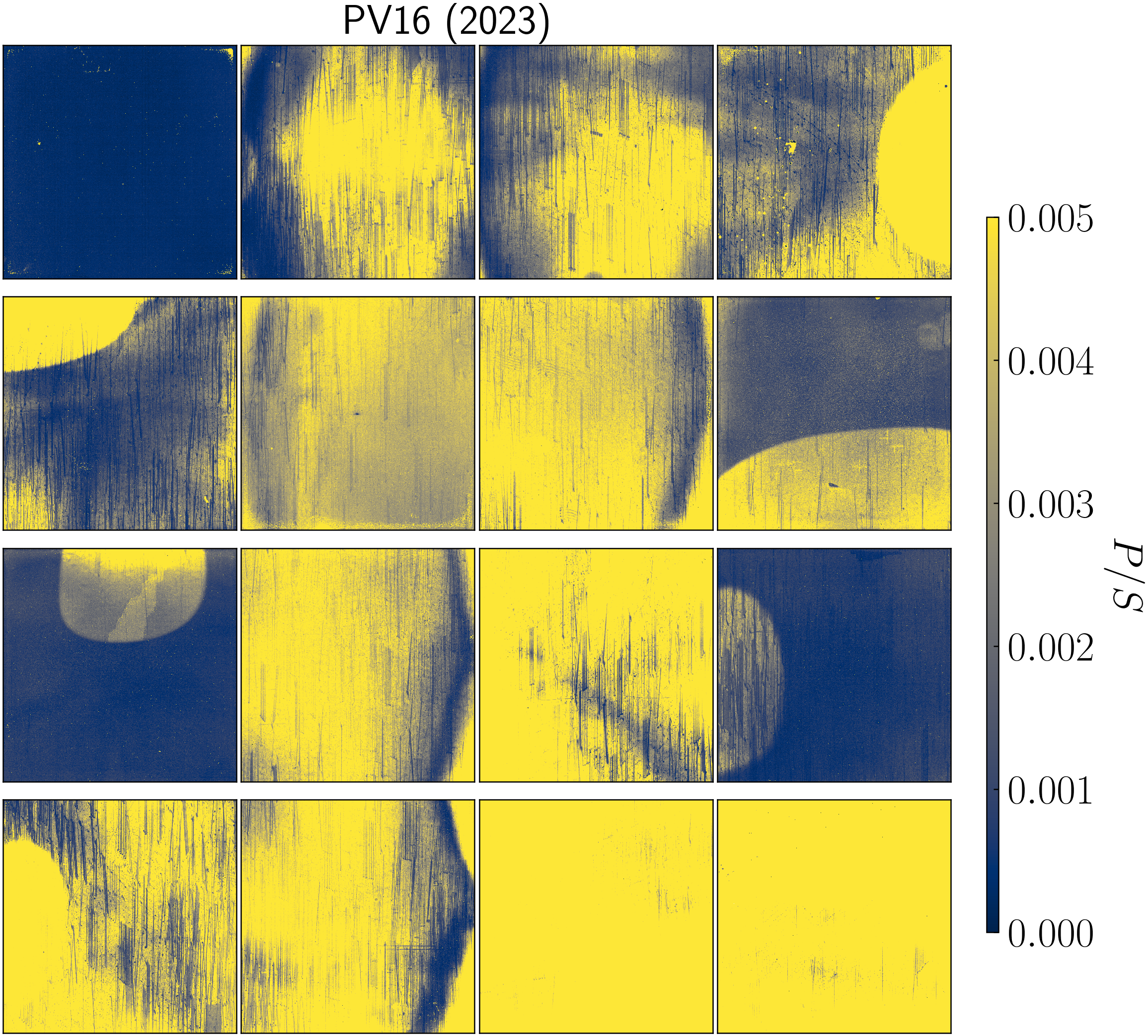}\qquad
    \includegraphics[width = 0.4\linewidth]{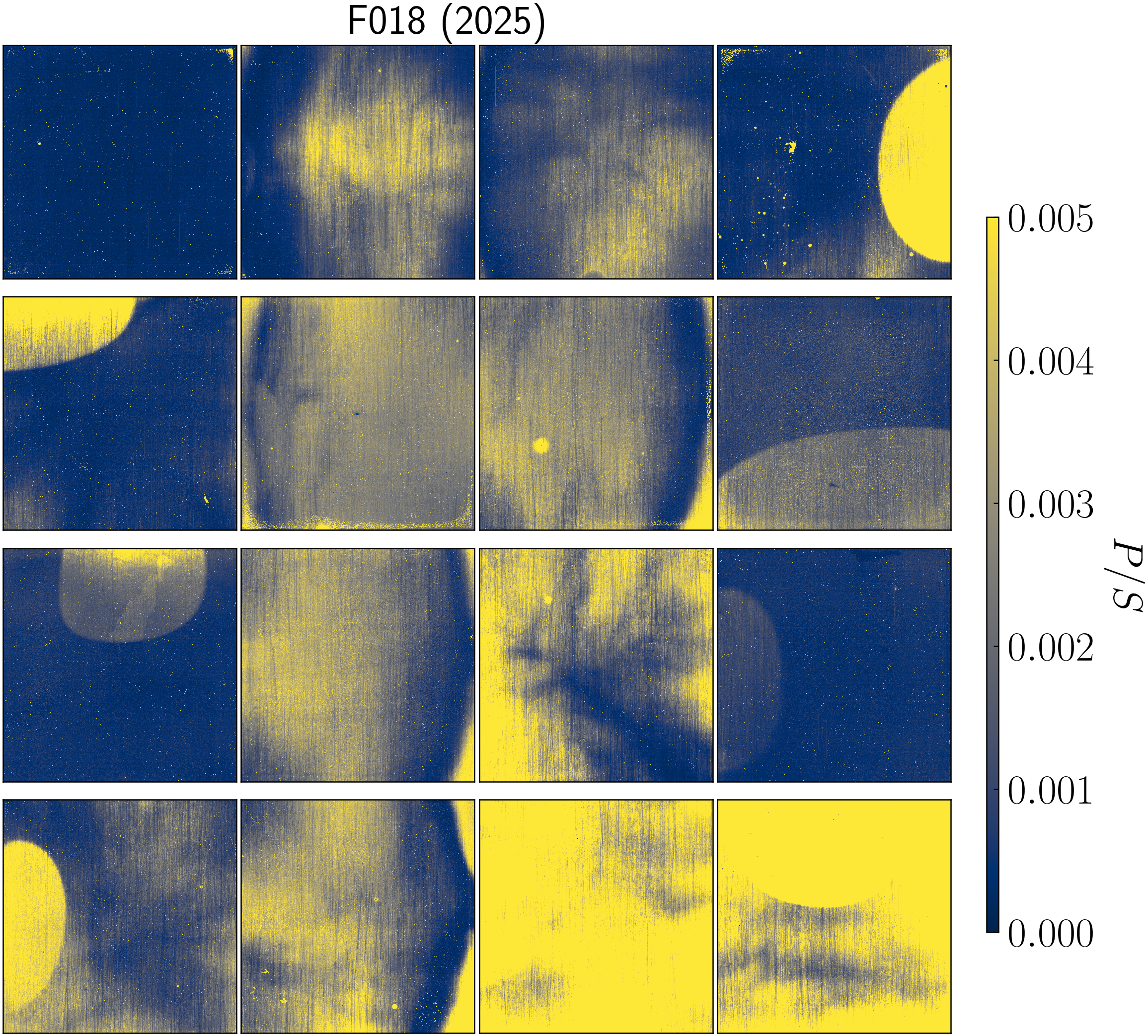}
    \caption{Persistence structures across detectors following a flat-field exposure of approximately $60\,$ke$^{-}$ for DCL (top left panel), TV1 (top right panel), PV-016 (bottom left panel) and F-018 (bottom right panel) data sets. Each detector has a size of $2048\times2048$ photosensitive pixels. $P/S$ is the persistence signal $P$ accumulated over the photometric readout time $(87.2\,\rm{sec})$ divided by the integrated fluence $S$.}
    \label{fig:persistence_structures}
\end{figure*}

The persistence response to a given signal $S$ is highly inhomogeneous, varying across detectors and even between individual pixels within a single detector, and spanning several orders of magnitude. The spatial distribution of persistence across the detector arrays is illustrated in Fig. \ref{fig:persistence_structures}. In most of the detectors, persistence is not spatially uniform. Several detectors exhibit highly contrasted spatial patterns, with areas showing high levels of persistence relative to the rest of the array. These regions frequently display a semi-circular morphology, which is especially prominent in detectors 11, 14, 21, 24, 31, 34, or 44. The remaining detectors show irregular spatial structures, while detector 41 appears largely homogeneous.

\begin{figure}
    \centering
    \includegraphics[width=\linewidth]{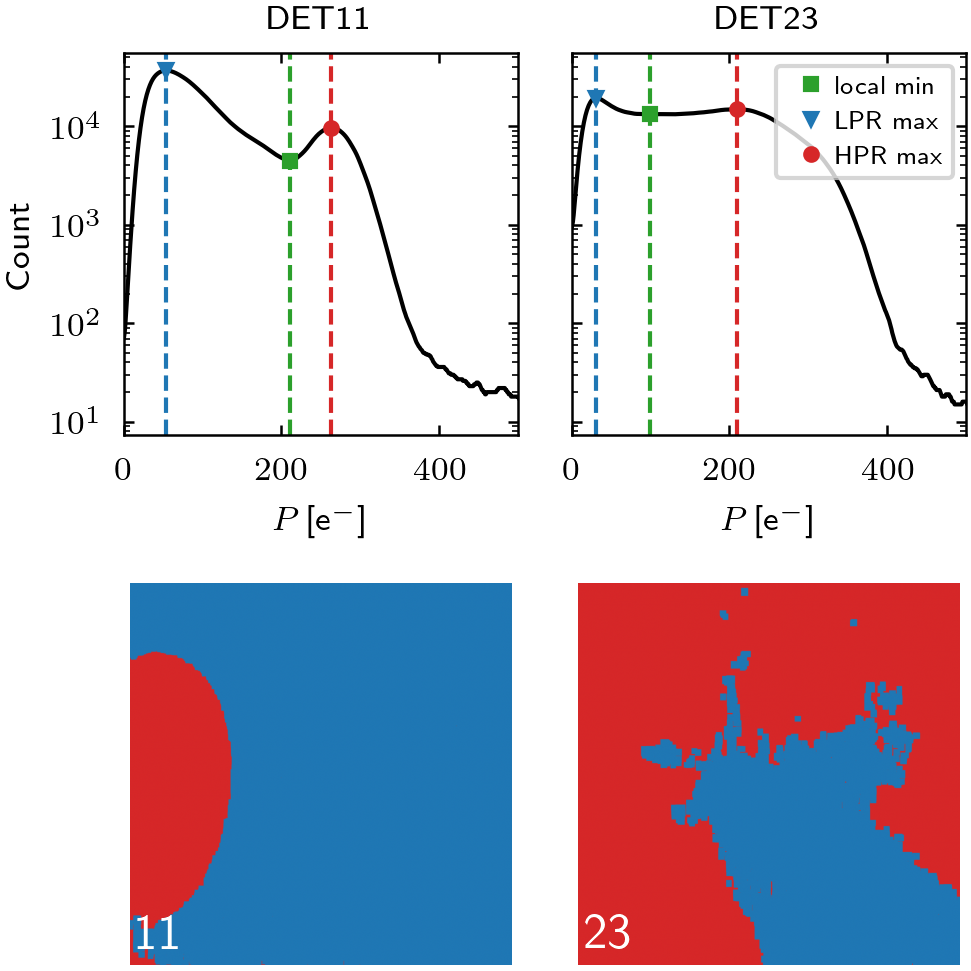}
    \caption{\textit{Top panels:} Distribution of persistence amplitudes used to identify HPR and LPR in the detectors in the F-018 data set. \textit{Bottom panels:} Spatial distributions of intra-detector regions. HPRs are shown in red, LPRs in blue. A sharp semicircular HPR is seen in DET11, while irregular regions are seen in DET23.}
    \label{fig:hists}
\end{figure}

To analyze persistence within each detector, we bin the amplitude of the first dark exposure acquired after a flat-field of 60\,ke$^-$ in the F-018 data set into a histrogram, as shown in the upper panel of Fig. \ref{fig:hists}, and smooth with a gaussian filter to reduce noise. In many detectors the histrogram has a well defined minimum between the two peaks which define a low persistence region (LPR) and a high persistence region (HPR). Examples of the spatially distribution of the LPR and HPR are shown in the lower panels of Fig. \ref{fig:hists}.

A region is not necessarily spatially connected. Since regions are defined by similarity in persistence amplitude rather than geometric continuity, they may have irregular shapes and consist of multiple disconnected components. Persistence is measured, modelled and applied to the flight data for each pixel. Pixels within the LPR or HPR behave similarly, so for clarity of presentation, the median persistence values are used for the following discussion.

\subsection{Signal dependence}
\label{sec:signal_dependence}

\subsubsection{Persistence amplitude -- signal dependence}
\label{sec:amplitudes_signal_dependence}

The signal dependence of the initial persistence amplitude is illustrated in Fig.~\ref{fig:alpha_fpa}, where the initial persistence amplitudes $\alpha$ and $\tilde{\alpha}$, derived from ground and in-flight calibration data sets, TV1 and PV-016, respectively, are shown as a function of the preceding fluence level. This dependence, for non-saturating sources, can be described as
\begin{equation}
    \alpha(S) = \left( a_\alpha + b_\alpha 10^{-5} S \right)
    \left[ 1 - \exp\left( -\frac{10^{-5} S}{c_\alpha} \right) \right]\,,
\label{eq:alpha_S}
\end{equation}
where $a_\alpha$, $b_\alpha$, and $c_\alpha$ are empirical parameters determined from calibration data and should not be assigned any physical interpretation. This formulation captures the nonlinear increase of the initial persistence amplitude with increasing fluence. Fits of Eq.~(\ref{eq:alpha_S}) to below-saturation TV1 and PV-016 data sets are also shown. The best-fit parameter values are strongly correlated and are provided for reference only.

A direct quantitative comparison between $\alpha$ and $\tilde{\alpha}$ is not straightforward because of differences in the data acquisition schemes and processing procedures (see Sect.~\ref{sec:parameterisation}). Nevertheless, despite these formal and methodological differences, both data sets exhibit qualitatively similar trends in the dependence of the initial persistence amplitude on the preceding illumination. This increase of the initial persistence amplitude with the preceding stimulus $S$ is commonly interpreted as a consequence of the expanding depletion region exposing a growing number of traps as the signal increases \citep{Smith_2008a,Long_2013}. The observed nonlinearity of the increase may reflect a lower trap density near the boundary of the depleted volume, a nonlinear dependence of depletion depth on fluence, or a combination of both effects.

\subsubsection{Persistence decay rate -- signal dependence}
\label{sec:decay_signal_dependence}

The signal dependence of the persistence decay rate is illustrated in Fig.~\ref{fig:beta_fpa}, where the decay indices $\beta$ and $\tilde{\beta}$, derived from ground and in-flight calibration data sets (TV1 and PV-016, respectively) are shown as a function of the preceding fluence level. In both cases, the data exhibit the `brighter–faster' effect, whereby persistence induced by higher fluence decays more rapidly than persistence generated by weaker stimuli. This dependence can be described by 
\begin{equation}
    \beta(S) = a_\beta \left(1 + \frac{S}{b_\beta} \right)^{c_\beta}\,,
\label{eq:beta_S}
\end{equation}
where $a_\beta$, $b_\beta$, and $c_\beta$ are empirical parameters and, as in the case of $a_\alpha$, $b_\alpha$, and $c_\alpha$, should not be interpreted as having any physical significance. Although a direct quantitative comparison between $\beta$ and $\tilde{\beta}$ is complicated (see Sect.~\ref{sec:parameterisation}), the values of the decay indices derived from the two data sets show substantially better agreement than the corresponding initial amplitudes. The variation of the decay index with source fluences ranging from 20 to 80\,ke$^-$ is approximately 13\% in HPRs and 35\% in LPRs. On a detector-by-detector basis, this variation ranges from about 10\% to 30\% in HPRs and from 15\% to 45\% in LPRs. 

In addition to the signal dependence, systematic differences are observed between HPR and LPR. For a given stimulus amplitude, the decay slopes in HPRs are generally steeper than those in LPRs. Since the preceding fluence level is the same in both regions, this difference cannot be attributed to the brighter–faster effect. Instead, it likely reflects intrinsic differences in the persistence properties of the detector material, possibly related to variations in the trap density.

While the dependence of the initial persistence amplitude on the preceding fluence is intuitive, the dependence of the decay index $\beta$ on stimulus amplitude is less straightforward. Although our results show an increase of $\beta$ with stimulus flux, corresponding to a steeper decay at higher fluence, earlier studies report an approximately invariant decay shape with stimulus amplitude for unsaturated exposures \citep{Smith_2008b}. Moreover, because $\beta \le 1$ implies a divergent total released charge as $t \to \infty$, the adopted parameterisation must be interpreted as valid only over a finite time interval.

\subsection{Differences between ground tests and in flight characterisation}
\label{sec:time_evolution}

\subsubsection{Persistence amplitudes -- time evolution}
\label{sec:amplitudes_time_evolution}

\begin{figure}
    \centering
    \includegraphics[width=\linewidth]{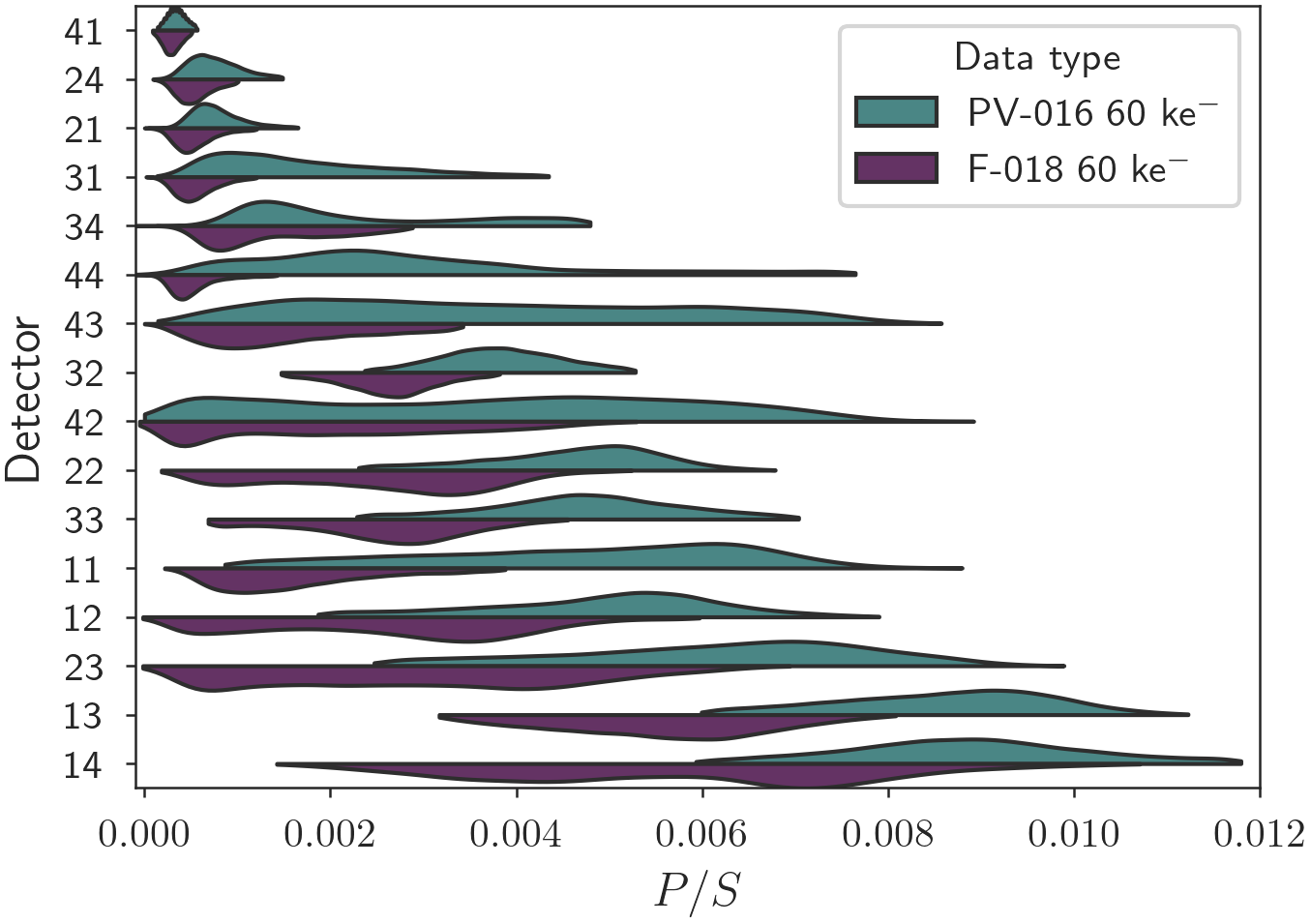}
    \caption{Per pixel distributions of $P/S$ after a prior fluence $S$ of 60\,ke$^-$ in flat-field illumination measured in PV-016 (2023) and in F-018 (2025).}
    \label{fig:persistence_evolution_violin}
\end{figure}

Figure \ref{fig:persistence_structures} shows snapshots of the persistence signal measured at different stages of the mission, comparing ground-based calibration data with in-flight observations. In addition to the TV1 data set, the figure includes a snapshot obtained during acceptance testing at Detector Characterisation Laboratory (DCL) at NASA/GSFC in 2017\footnote{The DCL acceptance-test data set, acquired in 2017, combines measurements from six independent tests, each involving four detectors. The measurements were performed at an operating temperature of approximately 100\,K. A fluence of 60\,ke$^{-}$ was delivered over 560\,s during these tests. The subsequent dark exposures were sampled at 60\,s intervals over a total duration of 10 minutes. To ensure a more consistent comparison with the TV1 and in-flight datasets, only the first three dark frames were used to estimate the persistence current over a comparable exposure time. }. The corresponding operating temperatures were approximately 100\,K for the DCL measurements, 87\,K for TV1, and 95\,K for the in-flight measurements (PV-016 and F-018 data sets).

A direct quantitative comparison between the DCL data set and the other data sets is not straightforward because of differences in the data acquisition procedures. In particular, the illumination timescale differs significantly: a fluence of 60\,ke$^{-}$ was delivered over 560\,s during DCL testing, compared with 110\,s for the TV1 and in-flight measurements. The temporal sampling of the subsequent dark exposures also differs, with frames acquired at 60\,s intervals for DCL, compared with a cadence of 1.45\,s for the other data sets. Nevertheless, the comparison provides useful insight into the evolution of the persistence behaviour. For comparable flat-field illumination levels of approximately 60\,ke$^{-}$, the persistence amplitudes measured during TV1 are systematically lower than those observed during the PV phase in September 2023, at the beginning of the survey. This difference can be attributed, at least in part, to the different operating temperatures. The persistence in PV-016 data set is more consistent with the results obtained during acceptance tests at DCL -- both data sets were collected at temperatures of approximately 95 and 100\,K, respectively, while the TV1 measurement was performed at a temperature of 87\,K. This clearly indicates that the operating temperature plays a significant role in determining the amplitude of persistence.

The comparison of the two bottom panels of Fig. \ref{fig:persistence_structures} further indicates that the persistence level decreased over the course of the mission. In the F-018 calibration data set (March 2025), the persistence response following a flat-field exposure comparable to that used in PV-016 is reduced and approaches the levels measured during ground calibrations. At the same time, the mean flat-field signal differs by approximately 3\% between PV-016 and F-018. Figure~\ref{fig:persistence_evolution_violin} presents the distributions of the ratio $P/S$ measured after a prior fluence of $S = 60\,\mathrm{ke}^{-}$ under flat-field illumination for PV-016 (2023) and F-018 (2025). On average, we observe a reduction in $P/S$ of approximately 37\% in the HPRs and 59\% in the LPRs between the PV-016 and F-018 measurements. The decrease varies across detectors, ranging from 25\% to 55\% in the HPRs and from 55\% to 80\% in the LPRs.

\begin{figure*}
    \centering
    \includegraphics[width=\linewidth]{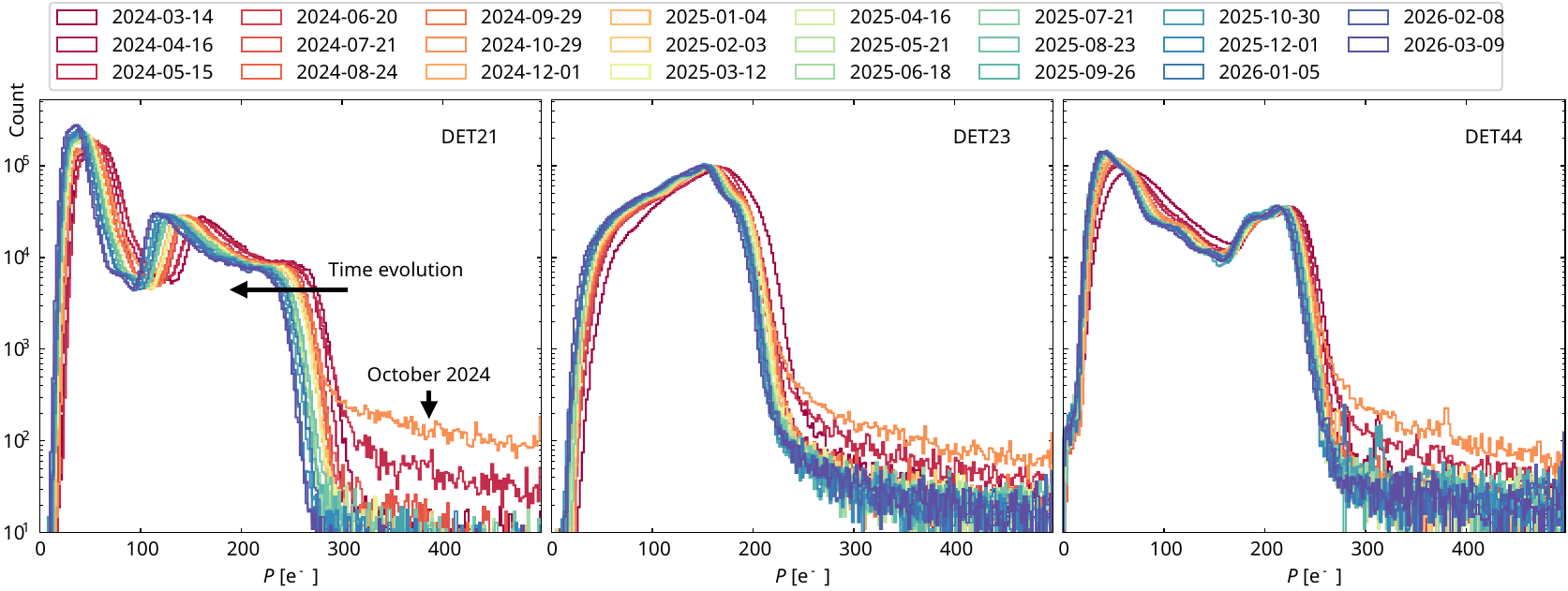}
    \caption{Distribution of the persistence response in the monthly F-007 data sets consisting of dark frames acquired after a series of flat-field exposures at 8\,ke$^-$ (\SI{90}{e^{-}.s^{-1}}) in photometric readout mode. The enhanced high-persistence tail visible around October 2024 correlates with an increased rate of EPHs in those data.}
    \label{fig:persistence_evolution_hist}
\end{figure*}

To investigate whether the transition between 2023 and 2025 is smooth or abrupt, we analysed the F-007 calibration data, consisting of dark frames acquired after a series of flat-field exposures at 8\,ke$^-$ (\SI{90}{e^{-}.s^{-1}} in photometric readout mode). Although these observations were not specifically designed to monitor persistence, they provide a useful proxy for tracking its temporal evolution. Figure \ref{fig:persistence_evolution_hist} presents the distribution of the persistence response in the monthly F-007 data sets. A clear decrease with time is observed, indicating that the transition between 2023 and 2025 is gradual rather than abrupt. Figure \ref{fig:evolution_flight_fpa} shows the median persistence signal computed for both HPRs and LPRs normalised to the first observation in March 2024. The figure also includes the temporal evolution of the median flat-field variation over the same period. In contrast to the flat-field signal, which varies by only about 3\% over the same time range and shows no appreciable difference between HPRs and LPRs, the initial persistence amplitude exhibits a substantially stronger dependence with time. On average, it decreases by approximately 10\% in HPRs and 25\% in LPRs. For individual detectors, the observed variation spans 3--25\% in HPRs and 12--43\% in LPRs. This differential behaviour complicates the interpretation of the temporal evolution and adds additional challenges for modelling persistence. A direct quantitative comparison of the persistence measured in PV-016 (and F-018 during the nominal mission) with that measured in F-007 is, however, not straightforward, since the data sets were acquired using different fluences and the persistence response depends on the applied stimulus. Nevertheless, both F-018 and F-007 independently show a decrease in persistence over time, although at different rates.

\begin{figure}
    \centering
    \includegraphics[width=\linewidth]{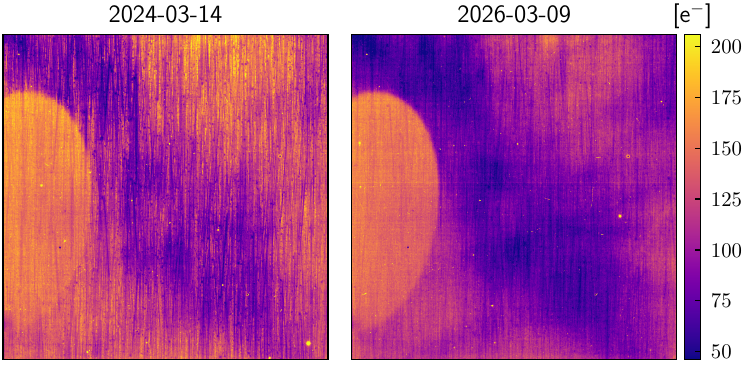}
    \caption{First dark frame following a \SI{90}{e^{-}.s^{-1}} flat (F-007) for DET11. \emph{Left}: March 2024. \emph{Right}: March 2026. In the earlier data, narrow elongated structures are visible in the persistence signal, showing significantly lower amplitude than the surrounding pixels. In the later data, broader diffuse regions of reduced persistence appear. The semicircular region of enhanced persistence remains visible in both epochs.}
    \label{fig:stars_saturation}
\end{figure}

Despite temporal variations in amplitude, the spatial morphology of the persistence regions remains remarkably stable between ground and in-flight measurements. Only small-scale differences are observed, as illustrated in Fig. \ref{fig:stars_saturation}. In particular, the F-007 data set (March 2024) reveals narrow, elongated regions of reduced persistence, which were not observed during ground characterisation. Similar structures are already visible in the PV-016 data set (2023), indicating that they did not appear exclusively after the beginning of the nominal mission. However, we cannot determine precisely when these structures first appeared or what caused their evolution between TV1 and PV-016. Furthermore, these structures are not stable over time. Similar structures are also present in the F-018 (2025) data set, although not necessarily at the same detector locations. The comparison between PV-016 and F-018 data sets indicates that some traces disappear while others reappear, and their overall number increases. As the mission progresses, these narrow, elongated regions also become increasingly diffuse and extend over a larger fraction of the detector. These features are likely linked to prior saturation events. Given their orientation, the elongated traces of reduced persistence are most probably associated to previous spectroscopic exposures that drove pixels into oversaturation. The diffuse regions of reduced persistence may result from repeated saturation episodes, such as those induced by bright stars, which could progressively modify the persistence response. Although this interpretation remains speculative, it is consistent with trends observed during ground characterisation, which indicate that oversaturation can reduce the persistence response over time (see Sect. \ref{sec:post_saturation} for details).

\subsubsection{Persistence decay rate -- time evolution}
\label{sec:decay_time_evolution}

The modification of the persistence decay index over time is shown in Fig.~\ref{fig:decay_flight_fpa}. The difference between the decay index $\tilde{\beta}$ derived from data sets acquired in 2023 (PV-016) and 2025 (F-018), following a 60\,ke$^-$ flat-field, is below 5\%. A slightly larger variation is observed in LPRs (around $5\%$) than in HPRs (around $2\%$). Overall, the decay indices $\tilde{\beta}$ measured in 2025 are systematically lower than those obtained in 2023, indicating a slower decay. The difference between the ground-based decay index $\beta$ and the in-flight value $\tilde{\beta}$, is more likely attributable to differences in data sampling (as discussed in Sects.~\ref{sec:parameterisation} and \ref{sec:decay_signal_dependence}) rather than to a temporal evolution of the persistence properties.

\subsection{Persistence after saturation}
\label{sec:saturation}

\subsubsection{Persistence of over-saturation signals}
\label{sec:over_saturation}

The persistence response of the detectors to highly saturating stimuli was partially characterised during ground calibrations and can also be inferred indirectly from in-flight data. These observations indicate that the persistence behaviour above saturation differs significantly from that observed for sub-saturation stimuli.

\begin{figure}
    \centering
    \includegraphics[width=\linewidth]{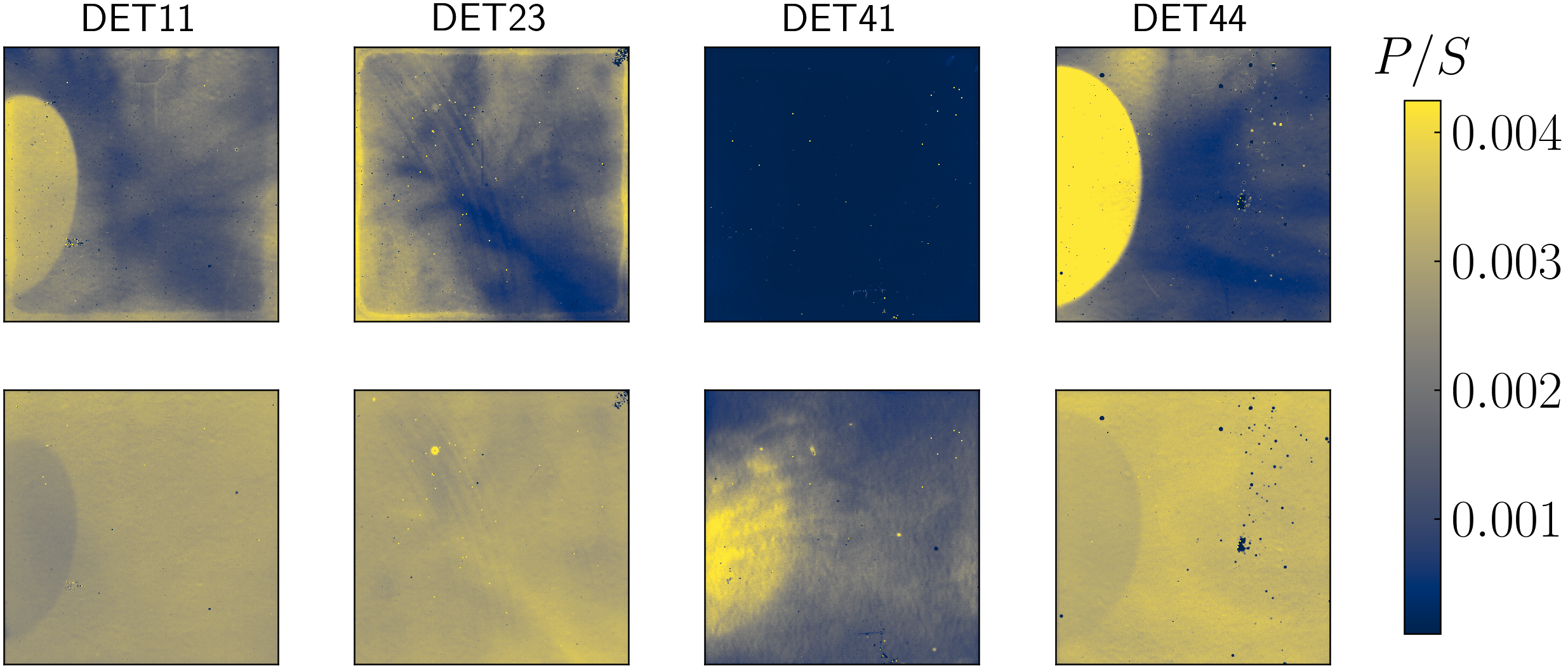}
    \caption{Persistence after a stimulus of 100 ke$^{-}$ -- close to but below pixel saturation (top panel) and after a stimulus 280 ke$^{-}$ -- well above the pixel saturation (bottom panel). This is for the TV1 data set.}
    \label{fig:pers_sat}
\end{figure}

One of the most notable effects is a reversal of the relative persistence amplitudes between the HPR and LPR compared to the behaviour observed for sub-saturation fluence. This effect is illustrated in Fig. \ref{fig:pers_sat}, which shows persistence maps following a stimulus close to, but below, saturation (top) and after a stimulus exceeding the saturation level (bottom). For several detectors, regions that exhibit higher persistence than the rest of the array after sub-saturation flat-field exposures show the opposite behaviour after highly saturating fluence. In these regions, the persistence signal becomes lower than in the surrounding pixels, effectively reversing the contrast pattern. In short, LPRs become HPRs and HPRs become LPRs. Detector 41 exhibits a different trend: for sub-saturation stimuli it shows relatively low and spatially homogeneous persistence, whereas after a highly saturating flat-field a distinct semicircular region of enhanced persistence emerges. 

The evolution of the persistence response with increasing flat-field levels, including over-saturating fluences, is shown in Fig.~\ref{fig:saturation_mosaic}. Each detector follows a different relation and no standard behaviour is observed. In some low-persistence detectors (interpreted as detectors with less active traps) little evolution of $P/S$ is observed below saturation, while a strong increase in persistence is observed above saturation. Traps have not been `saturated', rather the number of accessible or active traps has increased. In high-persistence detectors, with many active traps, the ratio $P/S$ continues to decrease for fluxes above saturation, as if there were a saturation effect.

Furthermore, the initial persistence amplitude, $\alpha$, and the decay index, $\beta$, measured following highly saturating stimuli deviate from the empirical relations derived for sub-saturation flat-field exposures (Eqs.~\ref{eq:alpha_S} and \ref{eq:beta_S}, respectively), as illustrated in Figs.~\ref{fig:alpha_fpa} and \ref{fig:beta_fpa}. In most cases, $\alpha$ after over-saturation exceeds the value predicted by extrapolation of the low-fluence model. In addition, the contrast between LPRs and HPRs is significantly reduced, although the amplitudes do not converge towards a common asymptotic value with increasing fluence. On the other hand, the decay index after over-saturation becomes shallower than predicted for HPRs and steeper than predicted for LPRs, leading to a reduced difference in the power-law index between the regions.

These results indicate that the persistence response after fluences above saturation differs fundamentally from that observed at lower fluence levels, suggesting that additional physical processes contribute to persistence when the detector is driven beyond its full-well capacity. At present, we are unable to identify the underlying cause of this behaviour.

\subsubsection{Long term impact of over-saturation signals on persistence response}
\label{sec:post_saturation}

\begin{figure}
    \centering
    \includegraphics[width=\linewidth]{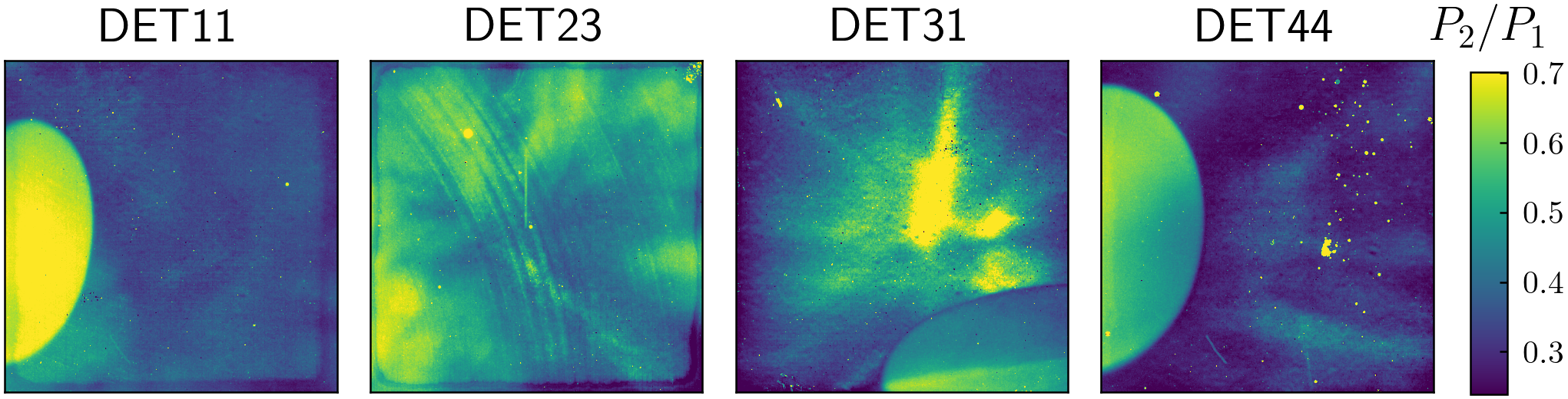}
    \caption{\label{fig:persistence_after_saturation}
    Ratio of initial persistence amplitudes $P_2/P_1$ measured after a fluence of 70\,ke$^{-}$ in flat-field illumination (TV1 data set). The first measurement, $P_1$, was made before any saturation -- detectors were under various illuminations but always below saturation for a week after cooling down. The second measurement, $P_2$, was made 8 hours after an over-saturation to 400 ke$^{-}$.}
\end{figure}

A prior saturation event alters the subsequent persistence behaviour of the detectors, leading to a reduced persistence response that remains observable over timescales of several days. This effect is illustrated in Fig.~\ref{fig:persistence_after_saturation}, which compares persistence measured after the same 70\,ke$^-$ flat-field stimulus under two conditions during ground tests. The first measurement, $P_1$, was obtained before any saturation event, following a week of sub-saturation fluence after cooldown. The second measurement, $P_2$, was acquired 8 hours after an over-saturation to 400\,ke$^-$. Eight hours after the saturation event, the persistence amplitude is reduced by approximately 40\,\% on average.

Consistent trends are observed in the in-flight data, as commented in Sect. \ref{sec:amplitudes_time_evolution}. In the F-018 data set, persistence amplitudes are lower than those measured during the PV phase likely reflecting repeated exposure of pixels to saturation events during operations. These results indicate that over-saturation induces long-lasting changes in persistence, potentially lasting for months with little temporal evolution. Consequently, modelling persistence requires accounting for prior saturation history, or recalibrating persistence after oversaturations. 

\subsection{Impact of persistence on the stabilisation of flat-fields}
\label{sec:impact_on_flats}

When the fluence level changes -- either from higher to lower fluence or from lower to higher fluence -- the pixel response requires some time to reach a steady state, defined as the regime in which, under constant external illumination, the measured signal remains stable within the limits of statistical noise. In contrast, a non–steady-state regime corresponds to a situation in which, despite constant external flux, the measured signal systematically evolves with time, exhibiting either an increase or a decrease that cannot be attributed to statistical fluctuations.

To quantify the magnitude of this effect, we analysed PV-010 data in which the detectors were exposed to a sequence of constant fluence levels ranging from 8 to 64 ke$^-$ in flat-field illumination at several wavelengths using five LEDs (see Sect. \ref{sec:data_flight}). For each configuration, 30 photometric exposures were acquired over approximately one hour. We then evaluated the ratio $S_0/S_{\infty}$ between the initial signal $S_0$, measured immediately after a change in fluence level, and the corresponding steady-state signal $S_{\infty}$, computed as the average signal over the last ten exposures.

As shown in Fig.~\ref{fig:flat_LED_time_dep}, the detector response reaches a stable level, within 1\%, within at most 10 minutes, for all tested flux levels and LED wavelengths. The stabilization timescale may depend on the flux level, particularly during the first few minutes, but this dependence does not affect the overall 10-minute timescale for reaching a stable response. HPRs exhibit larger variations than LPRs, with a maximum initial difference of about 3\% between the initial and stabilised signals. The largest deviations in $S_0/S_{\infty}$, of order 3\%, are observed in detectors with the highest persistence (13, 14, and 44) within their HPRs, whereas in most cases the effect remains below 1\%. 

At the lowest flux level for each LED, the stabilisation behaviour differs, typically showing an initial excess signal relative to $S_{\infty}$. This occurs because the lowest flux level for a given LED follows the highest flux level of the previous LED in the PV-010 sequence. The exception is LED-A, which was measured after a series of dark exposures. In this case, the largest deviations correspond to an initial deficit of signal relative to $S_{\infty}$ during the first few exposures at the lowest flux level. 

Overall, the ratio $S_0/S_{\infty}$ exhibits only a weak dependence on the absolute signal level $S_{\infty}$, as shown in Fig.~\ref{fig:flat_LED_flux_dep}, and no dependence on the LED wavelength. If the stabilisation observed in the flat-fields is interpreted as a balance between carrier capture and release, the absence of wavelength dependence is consistent with the results reported by \citet{Rauscher_2014} for persistence and by \citet{Biesiadzinski_2011} for reciprocity failure. 

Ideally, reliable measurements of astronomical sources should be obtained only once the detector has reached a steady-state response. However, after a change in illumination from one flux level to another, the detector response requires approximately 10 minutes to stabilise, independently of the initial and final flux levels. This timescale is impractical during routine science observations. Consequently, history of the measured fluences should be taken into account in both masking and correction strategies. On the other hand, the flat-field series does not exhibit any clear evidence of persistence accumulation. The signal reaches a stable level rather than increasing over time, as would be anticipated if persistence contributions from previous exposures were cumulatively added to each subsequent flat-field exposure. This observation may indicate a different charge-release behaviour under dark and illuminated conditions.

\subsection{Persistence in science data}
\label{sec:pers_sience}

To characterise persistence under realistic flight conditions, we analysed time series extracted from science exposures obtained during routine \Euclid observations for DET13. From 43 observations in SkyPatch 38, corresponding to 172 spectroscopic exposures and 516 photometric images (172 for each filter), we extracted time series of 32 consecutive exposures for pixels that recorded a signal higher than 5000 electrons in the first spectroscopic exposure. This selection yielded $5\times10^6$ pixels (55\%) affected by spectra of stars or galaxies, $3.4\times10^6$ pixels (37\%) affected by EPHs identified by the on-board quality factor \citep{Kubik2016}, and $0.7\times10^6$ pixels (8\%) that were already saturated in the first exposure. The high fraction of EPHs detected in these exposures can be attributed to enhanced solar activity at the time of the observations. These time series were subsequently grouped into bins of increasing fluence measured in the first exposure. 

The full time series includes signals measured in both spectroscopic and photometric exposures, as they appear in the ROS. However, the two time series were analysed separately, since spectroscopic and photometric data are not directly comparable due to differences in exposure time and readout mode. 

\subsubsection{Persistence of standard objects}
\label{sec:stars}

\begin{figure*}
    \centering
    \includegraphics[width=\linewidth]{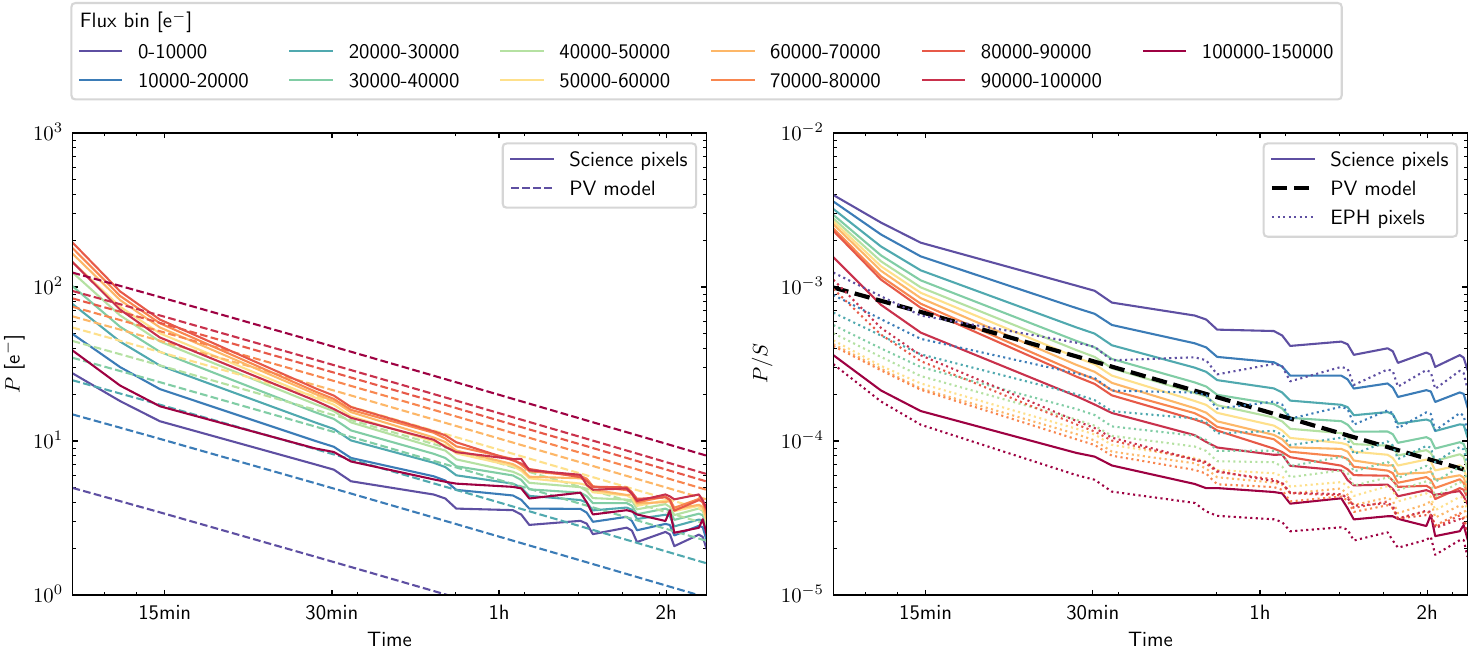}
    \caption{Persistence decays measured from time series extracted from science observations in DET13, compared with the model derived from PV-016 calibration data. \textit{Left}: Integrated persistence signal from science pixels in photometric targets compared to the model prediction. \textit{Right}: Ratio of the persistence signal to the preceding signal level for science pixels and EPHs, compared to the model prediction.}
    \label{fig:persistence_of_stars}
\end{figure*}

The left panel of Fig.~\ref{fig:persistence_of_stars} presents the binned time series for science pixels (i.e., pixels not affected by EPHs) compared with the model derived from PV-016 calibration data at 60\,ke$^-$. The differences between persistence measured during nominal science operations and that observed in dark calibration data are plausibly, although not exclusively, due to the presence of a non-negligible background during in-flight observations \citep{Scaramella-EP1}. This background includes both diffuse sky emission and residual persistence from previous exposures, which modify the trap occupancy and affect the subsequent persistence signal. This effect may be related to the differences in persistence behaviour observed under dark and illuminated conditions, as discussed in Sect. \ref{sec:impact_on_flats}. In addition, transitions between different background levels, arising from changes in exposure time between photometric and spectroscopic observations, can introduce additional charge trapping and release effects (see Sect. \ref{sec:impact_on_background}) that complicate the modelling of persistence decay. Furthermore, inter-pixel capacitance is not accounted for and may partially contribute to the observed discrepancies between the science and calibration data. However, its contribution is not expected to be significant. Finally, the persistence decay curves are binned in intervals of 10\,000\,ke$^{-}$ in stimulus signal $S$, which are relatively broad and therefore largely dominated by the dispersion in $S$, whereas the PV curves are averaged over the entire detector, where the $S$ dispersion is of order of a few percent. Both effects may contribute to the differences observed between the two data sets.

Despite the different observing conditions and the subtle effects that may affect the precise persistence values, we recover the typical qualitative behaviour of persistence observed in calibration data. The increase of the initial persistence amplitude with the preceding signal is clearly observed. The brighter–faster effect, whereby brighter sources exhibit faster decay, is also evident. 

\subsubsection{Persistence of EPHs}
\label{sec:cosmics}

In addition to astronomical sources, EPHs also produce measurable persistence signals. To study this, we extracted time series for $3.4\times10^6$ pixels affected by EPHs in the first exposure. EPH events in the science data are identified using the corresponding 8-bit QF value. Owing to the on-board processing performed by the NISP instrument, the exact occurrence time of an EPH, as well as the charge deposited by the event itself, cannot be recovered from the science data. Instead, the data provide only the total measured signal resulting from the ramp processing, which includes contributions from the EPH, the background, and, potentially, overlapping astronomical sources. In addition, the events identified as EPHs may occasionally include other types of anomalies. We therefore treat all events detected by the on-board processing as EPHs, without attempting to further discriminate between different types of events or anomalies.

As shown in the right panel of Fig. \ref{fig:persistence_of_stars}, the persistence amplitude associated with EPHs is typically lower than that measured for astronomical sources with comparable signal levels. In addition, the persistence decay following EPH events is typically faster than that observed for stars and galaxies. These differences confirm that persistence depends not only on signal amplitude but also on the temporal characteristics of the stimulus and the nature of the interacting particles. In particular, the soak time can significantly affect the persistence response. Unlike the flux from astrophysical sources, such as stars or galaxies, which can be assumed to remain approximately constant over the duration of an exposure, the timing of an EPH event varies from one exposure to another and may occur at any point during the integration ramp. Consequently, an EPH can produce a persistence response that differs from that induced by a constant illumination, owing to the different effective soak times experienced by the detector. Moreover, for stars and galaxies, each incident photon generates a single electron–hole pair in the HgCdTe substrate, with holes subsequently drifting through the substrate towards the cathode. In contrast, EPH produce a large number of ionisation electrons and holes distributed along a significant track length within the substrate, spanning a wide energy spectrum. As a result, the trapping probability may be different.

\subsubsection{Impact of persistence on the background estimate}
\label{sec:impact_on_background}

Large-scale persistence patterns are also evident in the background signal during nominal operations. The background level differs significantly between spectroscopic and photometric exposures because of different exposure times. In spectroscopic exposures, due to the longer exposure time of about 10 minutes, the accumulated background is of order of 1000\,e$^{-}$. In contrast, photometric exposures accumulate a background of roughly 100\,e$^{-}$ in about 1 min. As a result, during nominal observations the detectors are repeatedly transitioning between high- and low-background conditions throughout the observation sequence.

\begin{figure}
    \centering
    \includegraphics[width=\linewidth]{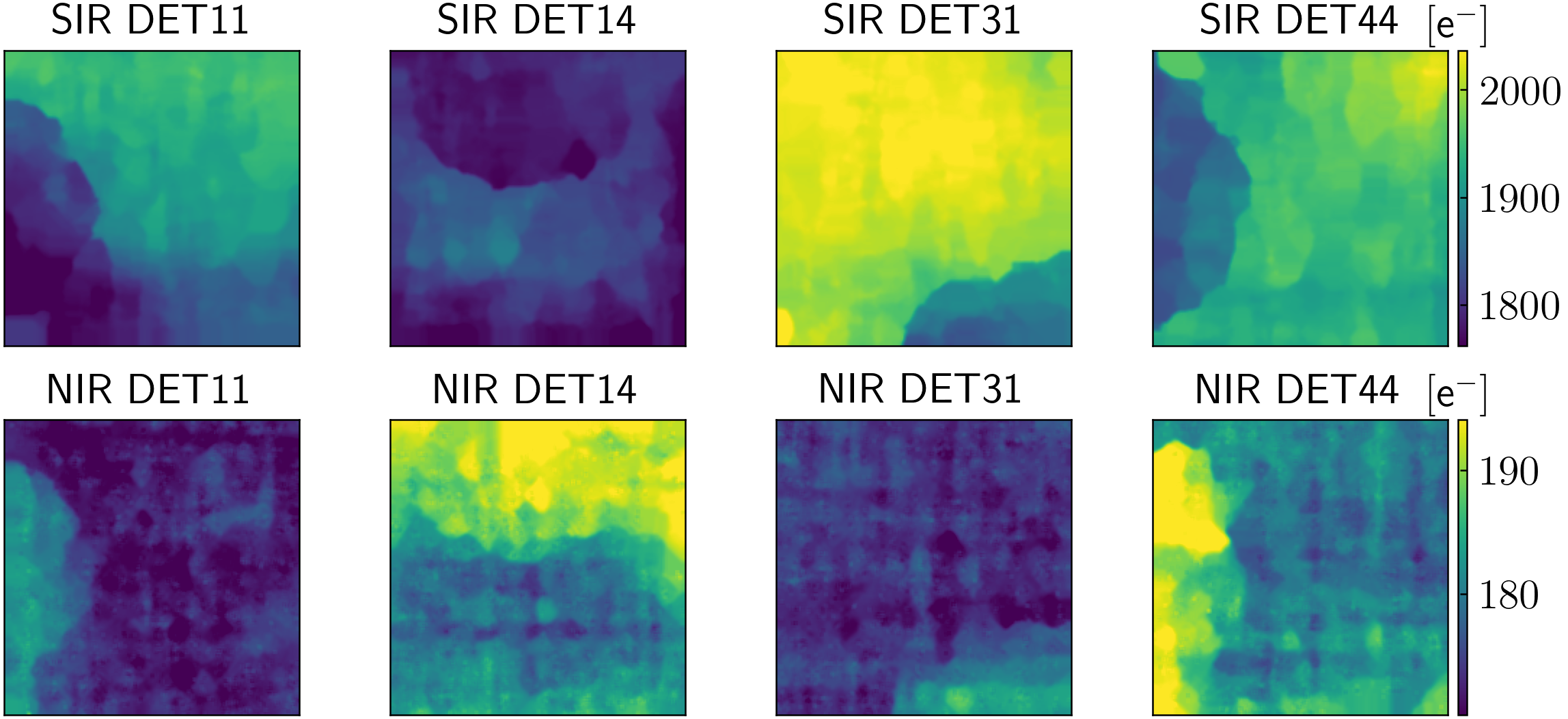}
    \caption{{\tt NoiseChisel} background  of spectroscopic (top panel) and photometric (bottom panel) exposures.}
    \label{fig:bkg}
\end{figure}

In spectroscopic exposures -- and particularly in the first spectroscopic exposure in each ROS which follows three photometric exposures and a slew dark -- the background measured with the {\tt NoiseChisel} software \citep{Akhlaghi2015} in HPRs is lower than in the rest of the image, as shown in the top panels of Fig. \ref{fig:bkg}. The difference is typically $\lesssim 1\%$. This behaviour is interpreted as charge trapping occurring after a period of low fluence (or dark conditions), which reduces the measured signal in these regions (similar to the flat-field stabilisation effect discussed in Sect. \ref{sec:impact_on_flats}). 

Conversely, in photometric exposures, especially in the \JE image that immediately follows a spectroscopic exposure, the background pattern in photometric exposures is similar to the persistence observed under dark conditions, following a transition from flat-field illumination to darkness. The background estimated in HPRs is systematically higher than in the rest of the image, as illustrated in the bottom panels of Fig. \ref{fig:bkg}, and the difference reaches up to 5\% in detectors exhibiting high contrast between HPRs and LPRs (DET11, DET14, and DET44), but is typically $\lesssim 2\%$. We attribute this excess to the release of charges trapped in the preceding spectroscopic exposure. 

\subsection{Power-law limitations}
\label{sec:limitations}

A simple power-law model does not fully capture the observed persistence decay. In the PV-016 calibration data, the first few points are typically underestimated by the fit (Fig.~\ref{fig:decay_flight_fpa}). Departures from a simple power-law are even more pronounced in the science data (Fig.~\ref{fig:persistence_of_stars}).

Two effects can account for these deviations. The first is related to the acquisition scheme used in the PV-016 data. The initial five dark exposures are obtained in photometric readout mode, whereas all subsequent exposures are acquired in spectroscopic readout mode, which has an exposure time approximately 5 times longer. This change in readout configuration can introduce a discontinuity in the time series and therefore represents an instrumental effect. However, analysis of ground characterisation data -- where all frames are available and different readout schemes can be applied a posteriori to the same decay sequence -- indicates that this effect is negligible at long times.

The second may be related to the presence of trap populations with a range of release timescales. For example, a combination of faster and slower charge-release processes could produce a decay that departs from a simple power law. However, the number of relevant trap populations and their characteristic timescales are not known a priori. In the science data, departures from a simple power-law are further enhanced by the presence of a non-zero background signal, which adds additional complexity to the modelling.

Several empirical approaches can improve the representation of the observed decay, but at the expense of additional model complexity. For example, modelling the decay with a finite sum of exponential components would introduce a large number of free and potentially correlated parameters, particularly when both amplitudes and time constants are fitted. Alternatively, the early and late stages of the decay could be fitted separately using different coefficients, or the decay could be approximated using higher-order polynomials in log-log space. While these approaches provide additional flexibility, they also introduce further free parameters and can increase the complexity and computational cost of the fitting procedure. Given the difficulty of constraining the detailed shape of the persistence decay, we therefore adopt a simple empirical parametrisation as a practical compromise between flexibility, robustness, and computational efficiency. This choice is intended as an operational parametrisation of the relevant slow component rather than as a precise physical description of the complete persistence decay.

Importantly, the fast component of persistence decay, which typically evolves over only a few tens of frames and is not included in our model, has a negligible impact on the masking procedure. For science operations, the key quantity is the timescale over which a pixel remains significantly affected by persistence. An accurate description of the slower decay component is therefore sufficient to determine when a pixel can be safely unmasked and considered suitable for subsequent science observations.

\section{Application}
\label{sec:application}

In standard ROS observations, persistence affects science exposures in several ways, each with different implications for data quality and analysis.

\paragraph{Spectro-to-photo persistence} occurs when residual charge from a preceding spectroscopic exposure is released during subsequent photometric exposures. This effect can lead to spurious source detections, contaminate object catalogues \citep{Q1-TP004}, and complicate the identification of spectra \citep{Q1-TP006}, since the spectroscopic extraction relies on accurate source positions derived from photometric detections. This effect has the largest impact both in terms of amplitude and the number of affected pixels, because spectroscopic exposures are typically longer, the corresponding fluences are generally higher. As a result, a substantially larger number of pixels reaches high signal levels, with approximately from 5 to 15\,\% of pixels typically exceeding a fluence of 1\,ke$^-$, a representative level comparable to the average background in spectroscopic exposures and used here as a reference for the applicability of the persistence model. These pixels can produce non-negligible persistence in subsequent photometric exposures.

\paragraph{Spectro-to-spectro persistence} arises when residual charge from a spectroscopic exposure affects subsequent spectroscopic exposures in another dither. This can contaminate spectra, potentially causing overlapping features and complicating spectra extraction \citep{Q1-TP006}. This effect is generally confined to exposures within the same ROS and is rare between different ROSs. The effect is weaker in the spectroscopic-to-spectroscopic than in the spectroscopic-to-photometric case. The interval between two spectroscopic exposures is longer, allowing the fast component of persistence to partially decay before the subsequent exposure. Although this reduction is partly offset by the longer integration time of the affected spectroscopic exposure, which increases the accumulated persistence charge, the higher background level in spectroscopic data compensates for this effect, resulting in a lower net impact.

\paragraph{Photo-to-spectro persistence} can generate artificial emission lines in extracted spectra when residual charge from photometric exposures contaminates subsequent spectroscopic observations. In terms of its overall impact, this effect is relatively minor. Typically, fewer than 0.5\,\% of pixels exhibit persistence signals exceeding 1\,ke$^-$. Although the longer integration time of the affected spectroscopic exposure increases the accumulated persistence charge, the higher background level in spectroscopic data largely compensates for this increase.

\paragraph{Photo-to-photo persistence} manifests in two ways. First, residual charge from a photometric exposure can affect subsequent photometric exposures in a different dither, producing spurious sources that are randomly distributed across the focal plane. Second, self-persistence occurs when the same source is observed multiple times within a single dither. Because the offsets between successive photometric exposures within a dither are small (a few pixels), a given source can illuminate the same pixels across the three exposures, thereby contaminating its own measured flux. In terms of the number of affected pixels, and excluding self-persistence, this is the least significant effect among all the considered cases. The interval between two photometric exposures in different dither positions is typically at least 15 min, allowing a substantial fraction of the persistence signal to decay. In addition, the target exposure time is relatively short and the number of high-signal pixels is small, further reducing the impact of the effect.

\paragraph{Large-scale persistence} is visible across the background of science exposures (see Sect. \ref{sec:impact_on_background}) and calibration products, such as flat-fields (see Sect. \ref{sec:impact_on_flats}). Moreover, it may impact measurements that rely on precise background characterisation, including nonlinearity calibrations and other detector response corrections.

\subsection{Persistence masking in \Euclid pipelines}
\label{sec:pipelines}

Persistence models derived from calibration data are used in \Euclid pipelines to detect and mask persistence in the science images. Q1 processing \citep{Q1-TP003}, used the persistence model derived from ground data characterisation using parameterisations given in Eq. (\ref{eq:I}) for the persistence current, and Eqs. (\ref{eq:alpha_S}) and (\ref{eq:beta_S}) describing the nonlinear dependence of the persistence amplitude and decay index on the previous fluence. This model included only the fast-decay component over short timescales. Only persistence from spectroscopic to photometric exposures within the same dither (15-minute history) was considered, while spectroscopic targets were not processed. 

For DR1 processing \citep{DR1-TP003}, the pipeline was refactored, and the persistence model updated using PV-016 calibration data. The adopted model is based on the decay of persistence following a 60\,ke$^-$ fluence in the PV-016 calibration data set, uses parameterisation given in Eq. (\ref{eq:Ipip}), while the fluence dependence of persistence parameters is neglected. The DR1 pipeline accounts for up to 5 hours of preceding exposures and includes spectroscopic targets. Masking of photometric targets was improved, and a Hough Transform-based method was introduced to enhance the detection and masking of persistence streaks in photometric images. Persistence in spectroscopic targets was also masked using the same decay model and the full photometry and spectroscopy history (see \citealp{DR1-TP003} for details).

These modifications result in a significant improvement in masking compared to Q1, since the longer illumination history and inclusion of both photometric and spectroscopic exposures allow faint persistence traces to be captured that were previously missed. However, the conservative pipeline configuration and the chosen masking parameters in DR1 sometimes lead to an excessive fraction of masked pixels. Further optimisation of the masking thresholds and temporal window is therefore required to reduce the fraction of masked pixels while preserving the effective suppression of persistence contamination, and is discussed in the following subsection. An even larger increase in the number of usable pixels could potentially be achieved through the application of persistence corrections. The performance of different correction approaches is investigated in Euclid Collaboration: Awad et al., in prep.

\subsection{Processing optimisation}
\label{sec:optimisation}

The performance of the persistence mitigation scheme, given the accuracy of the model, depends on both the adopted masking thresholds and the duration of the illumination history included in the persistence model. In DR1 processing \citep{DR1-TP003}, a 5-h time window was adopted to conservatively ensure masking of persistence from saturated sources. The masking threshold for photometric targets was set to 2\,e$^-$, following the current requirements. This value should not be interpreted as a physical detection threshold, as it is significantly below both the model precision (see Sect. \ref{sec:model_precision}) and the typical signal uncertainty. For a background level of 100 electrons -- the lowest signal that can be considered in photometric exposures -- the Poisson error is about 10 electrons. Thus, for the faintest sources exceeding the background, the uncertainty is even larger. Moreover, the threshold is independent of the underlying signal in the target exposure. A persistence signal of 2\,e$^-$ superimposed on a very bright object is treated in the same way as an identical persistence signal superimposed on a faint object. Together, these factors resulted in a significant proportion of masked pixels in photometric targets, sometimes exceeding 10\,\%. The threshold of 30\,e$^-$ applied to the spectroscopic target, keeps the fraction of masked pixels low, typically lower than 5\,\% per detector. Below, we discuss the model accuracy and investigate how the persistence-processing parameters, namely the history duration and the masking threshold, affect the fraction of masked pixels. 

\subsubsection{Model precision}
\label{sec:model_precision}

\begin{figure}
    \centering
    \includegraphics[width=\linewidth]{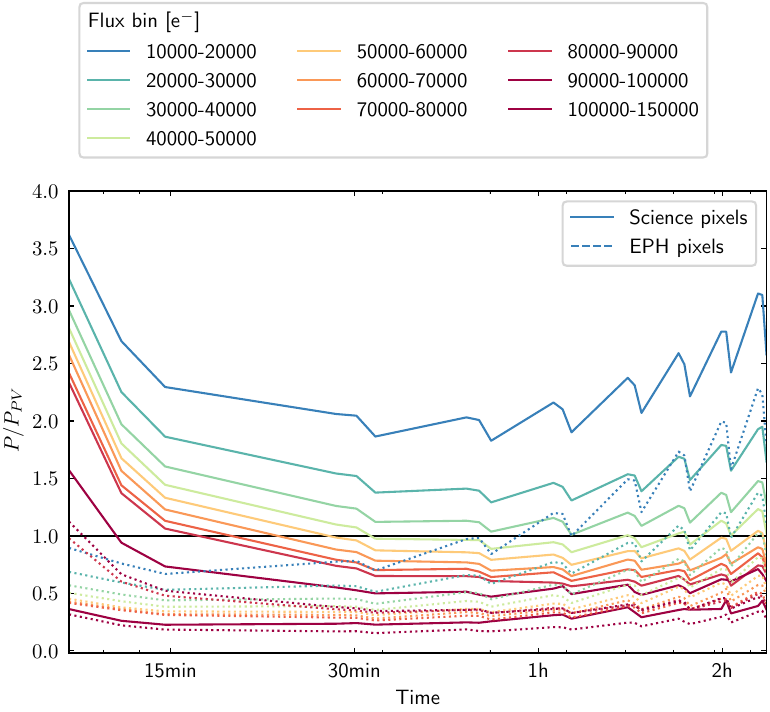}
    \caption{Ratio of the persistence signal measured in science exposures (SkyPatch 38) for stars and galaxies (solid lines) and EPHs (dotted lines) to the persistence model used in the \Euclid DR1 pipeline. The increase in the ratio at long times and low signals is caused by the persistence signal approaching the background level, while the model prediction continues to decrease.}
    \label{fig:model_residuals}
\end{figure}

The accuracy of the persistence model derived from the PV-016 calibration data and used in the \Euclid DR1 pipeline can be assessed by comparing the DR1 model predictions with the persistence signal of astrophysical sources measured above the background in science exposures, as illustrated in Fig. \ref{fig:model_residuals}.

For astrophysical sources, such as stars and galaxies, the model generally underpredicts the persistence signal, except at very high fluence levels and long elapsed times. For example, for DET13, in the first science exposure following illumination, the model underpredicts the persistence signal above the background by a factor of approximately 2.5 to 3.5 depending on the exact fluence level. Only for $S > 40\,000$\,e$^-$ and $t > 30\,$min does the model begin to overpredict the persistence signal. However, the time of transition from underprediction to overprediction depends on the deposited fluence. 

For EPH, the model generally overpredicts the persistence signal by a factor of about 2 on average. However, the exact level of disagreement depends on both the deposited charge and the elapsed time.

A quantitative assessment of the model accuracy is further complicated by the sensitivity of the measurements to the background estimate, which itself contains a persistence contribution, as discussed in Sect. \ref{sec:impact_on_background}, and by the temporal modification of persistence as discussed in Sect. \ref{sec:time_evolution}. A detailed evaluation of the model performance is therefore beyond the scope of this paper and is deferred to future work aimed at developing a more accurate description of persistence decay.

\subsubsection{Optimisation of the history time interval}
\label{sec:history_optimisation}

The time window used to select exposures in the illumination history determines how long persistence contributions are taken into account after a stimulus. To optimise this parameter, we used the persistence model derived from PV-016 calibration data to compute the decay of the persistence signal for each detector after a 100\,ke$^-$ flat-field illumination -- the highest achievable signal level below saturation.

\begin{figure}
    \centering
    \includegraphics[width=\linewidth]{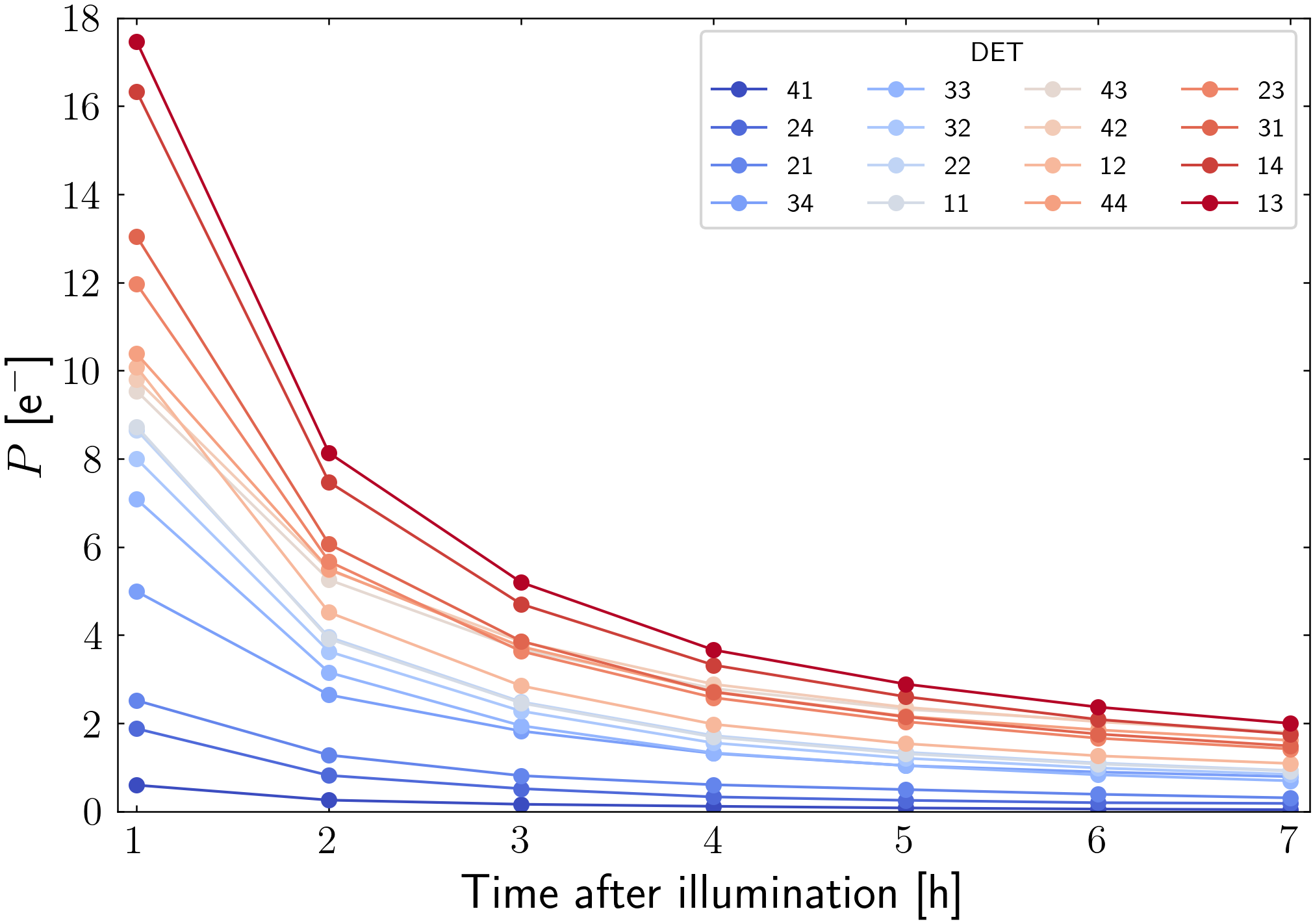}
    \caption{Median persistence signal predicted by the DR1 model, integrated over the photometric integration time, as  function of time after a fluence of 100\,ke$^-$ under flat-field illumination.}
    \label{fig:times_th}
\end{figure}

For the detectors exhibiting the strongest persistence, the time required for 50\% of pixels to fall below the nominal threshold of 2\,e$^-$ is of about 7\,h, as illustrated in Fig. \ref{fig:times_th}. Adopting higher thresholds leads to substantially shorter time scales: 50\% of pixels fall below 4\,e$^-$ after approximately 4\,h and below 10\,e$^-$ after les than 2\,h. The exact values, however, vary from detector to detector. Given (i) the broad distribution of persistence amplitudes across the focal plane, (ii) the fact that the 100\,ke$^-$ fluence represents the highest achievable signal level below saturation, reached in practice by only a tiny fraction of pixels, and (iii) the fact the at long time scales the model tends to overestimate persistence signal, a time interval of about 3\,h appears to provide a reasonable compromise between operational efficiency and mitigation of persistence effects.

\subsubsection{Optimisation of the masking threshold}
\label{sec:threshold_optimisation}

To optimise the masking threshold for photometric observations, persistence maps were generated using a 3-h illumination history. The fraction of masked pixels was then evaluated as a function of the threshold for a \JE exposure. The \JE  band was selected because it is expected to experience the strongest persistence contamination, since it follows directly after a spectroscopic observation. Two configurations were considered: (i) using only spectroscopic exposures in the illumination history; and (ii) including both spectroscopic and photometric exposures.

\begin{figure}
    \centering
    \includegraphics[width=\linewidth]{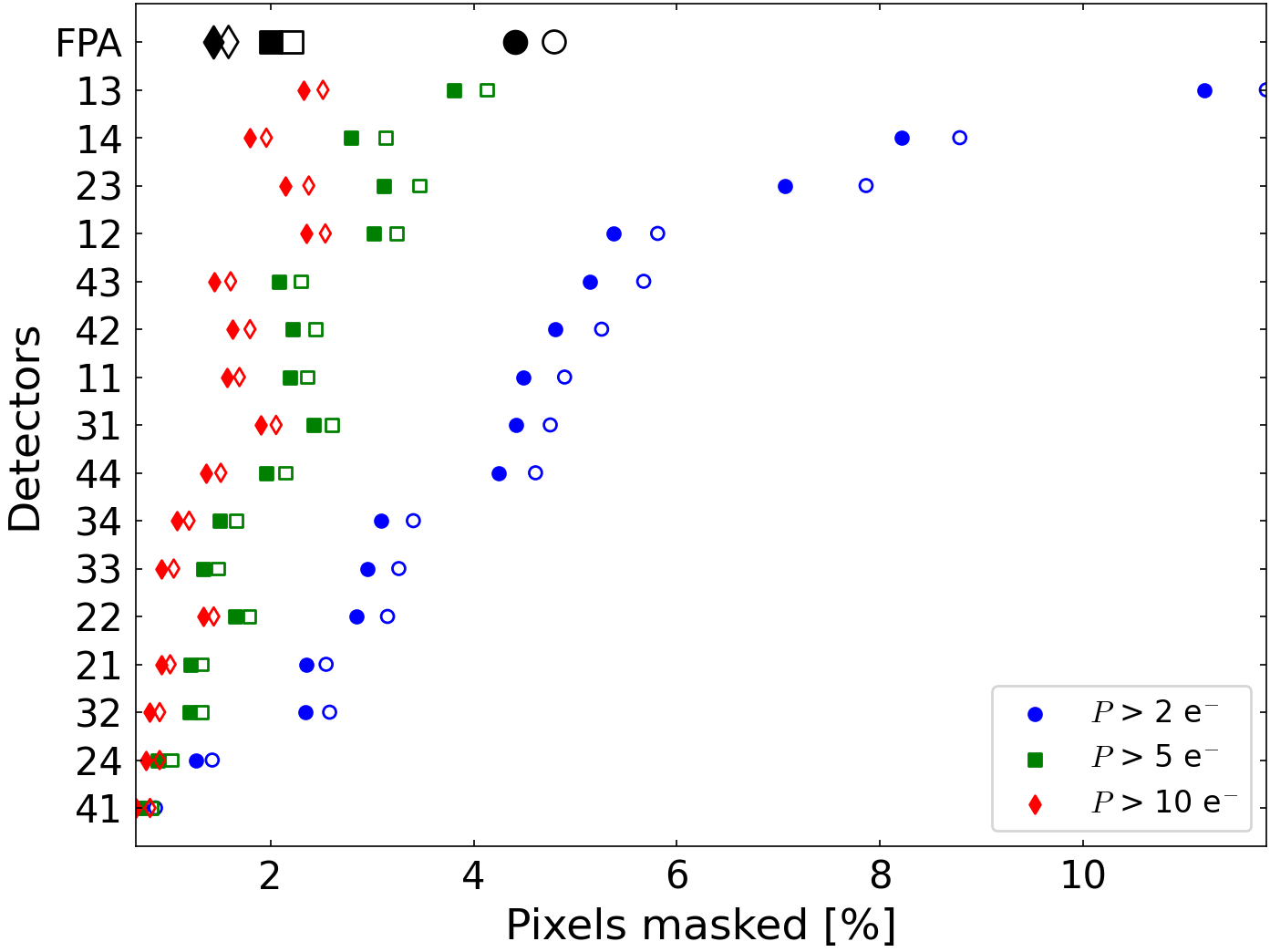}
    \caption{Fraction of pixels masked in a photometric \JE exposure for masking thresholds of 2, 5, and 10 e$^-$. Filled markers correspond to persistence images constructed using only spectroscopic exposures, while open markers include both spectroscopic and photometric exposures in the 3-hours history. For targets in the \HE and \YE bands, the fractions are slightly reduced due to a decrease in persistence amplitude caused by spectroscopic exposure, which is the dominant effect.}
    \label{fig:thresholds}
\end{figure}

\begin{figure}
    \centering
    \includegraphics[width=\linewidth]{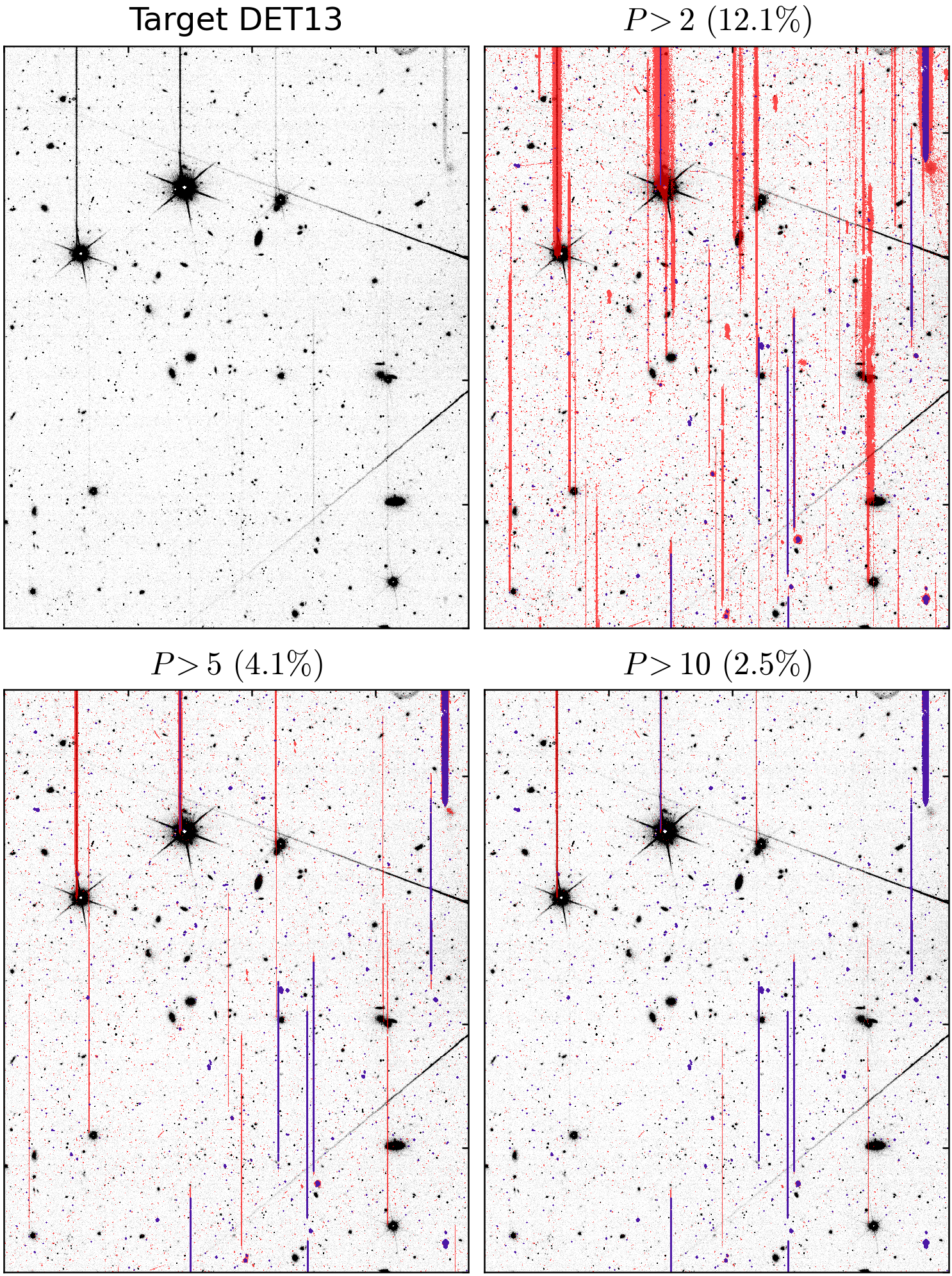}
    \caption{Persistence masking in the \JE photometric image for a $750 \times 940$-pixel region, shown for three persistence thresholds. Persistence image was computed using a 3-h illumination history, including both photometric and spectroscopic exposures. Masked pixels are shown in red, while pixels that saturated in any of the history exposures are shown in blue. The threshold value and the corresponding fraction of masked pixels (in parentheses) are indicated above each panel.}
    \label{fig:th_mask}
\end{figure}

As summarised in Fig. \ref{fig:thresholds}, for targets in the \JE band, a threshold of 2\,e$^-$ results in 1–10\% of masked pixels depending on the detector, corresponding to about 5\% on average for the full FPA. In the \HE and \YE bands, the percentages are slightly lower because the persistence signal has partially decayed since the preceding spectroscopic exposure. Increasing the threshold to 5\,e$^-$ decreases the masked fraction to $<$5\% for all detectors (about 2\% across the FPA), while still effectively removing visually identifiable persistence features, as shown in Fig. \ref{fig:th_mask}. A threshold of 10\,e$^-$ further reduces the masked fraction to $<$2\%, but leaves some persistence residuals unmasked. Including photometric exposures in the history increases the masked fraction by only 0.5\%. These results suggest that a threshold of 5\,e$^-$ may be close to optimal, although, a more robust optimisation based on a clearly defined objective function is still needed.

\section{Conclusions}
\label{sec:conclusions}

We have characterised and modelled detector persistence using complementary ground calibration data and in-flight measurements. Ground observations provide high temporal resolution and sensitivity to the fast component of persistence decay, while in-flight data enable the characterisation of the long-term evolution of persistence over timescales of up to about one hour. The different temporal regimes probed by these data sets lead to different parameterisations of the persistence models. 

Persistence is highly non-uniform across the focal plane, with variations of up to two orders of magnitude, both between detectors and within individual detectors. The persistence amplitude is $\sim$1\% of the preceding fluence, but has a modest nonlinearity trending lower persistence for higher fluence (when below pixel saturation). The decay rate also shows a dependence on fluence, with brighter sources providing faster decay. A single power-law model, Eq. \ref{eq:I}, is sufficient for masking pixels especially over long time scales of $\sim$5 hours which is sufficient to properly mask pixels as inoperable in the processing pipeline. However, the model does not properly predict decay on time scales within minutes after the initial stimuli.

Persistence measured in science observations seems to differ from that derived from dark calibration data. At low signal levels or short timescales, the model derived from calibration data in dark conditions tends to underpredict persistence from stars, while at longer timescales it typically overpredicts the signal. This difference is not yet fully understood. Moreover, EPH produce persistence signals with lower amplitudes and faster decay than astronomical sources with comparable signal levels, indicating that persistence depends not only on signal amplitude but also on the temporal characteristics and nature of the stimulus. This might be excepted because energetic particles generate a track of charge carriers well above the HgCdTe band gap on absorption compared to absorption of a single infrared photon which has an energy comparable to the HgCdTe band gap.

A gradual long-term decrease of persistence amplitude is observed during the mission, likely related to repeated exposure of pixels to highly saturating stimuli, as saturation produces lasting changes in persistence behaviour, reducing the persistence response and modifying the decay slope. Moreover, the persistence decay following highly saturating fluence deviates from empirical relations derived for sub-saturation conditions, suggesting that additional physical processes contribute when the detector exceeds full-well capacity.

Charge trapping and release affect all types of detector measurements. In typical calibration sequences of flat-fields, following transitions between different fluence levels, the detector response stabilises to within about 1\% within approximately 10 minutes. These effects also introduce small but measurable effects in the estimates of the background in science exposures. The spatial variation of the background amplitude, which can be attributed to charge trapping and release effects, is typically below a few percent, particularly in detectors exhibiting high persistence.

Persistence signal affects science observations in multiple exposure sequences (photo–photo, photo–spectro, spectro–photo, spectro–spectro), potentially introducing systematic effects in photometry and spectroscopy. Several of these effects are mitigated in the processing pipeline through masking, although self-persistence and large-scale background structures remain challenging. The effectiveness of persistence mitigation depends on both the adopted masking threshold and the length of the illumination history included in the model. The conservative configuration used in DR1 (5-h history, 2 e$^-$ threshold for photometric targets) ensures robust masking but results in a relatively high fraction of masked pixels. Tests indicate that a 3-h illumination history combined with a threshold of about 5 e$^-$ provides a good compromise, reducing the masked fraction to below about 5\% while still suppressing visible persistence artefacts. Nevertheless, a more robust optimisation of masking thresholds based on feedback from validation of science products, for example the spectral pipeline, is in work.

Future improvements to persistence modelling should focus on a more accurate description of persistence in scientific data, including self-persistence and the distinct persistence behaviour induced by EPHs. In addition, improved empirical or physically motivated models, together with systematic optimisation of masking parameters based on well-defined performance metrics, will be essential for maximising the scientific return of future \Euclid data releases.

Although the model presented in this work successfully identifies and masks pixels affected by persistence, accurately predicting the excess flux due to persistence remains challenging. Alternative correction methods are also being developed for potential inclusion in the \Euclid DR2 processing pipeline and later data releases. A comprehensive comparison of the proposed persistence correction techniques will be presented in a forthcoming study (Euclid Collaboration: Awad et al., in prep.).

\begin{acknowledgements}
    \AckEC
    \AckDRone
\end{acknowledgements}

\bibliographystyle{aa}
\bibliography{Euclid, main} 

\begin{appendix}
\onecolumn 

\begin{figure*}
\section{Persistence characteristics for the full focal plane}

In this appendix, we present the persistence characteristics discussed in the paper for each of the 16 detectors, providing the complete set of results that complements the analysis in the main text.

    \begin{center}
    \includegraphics[width=\linewidth]{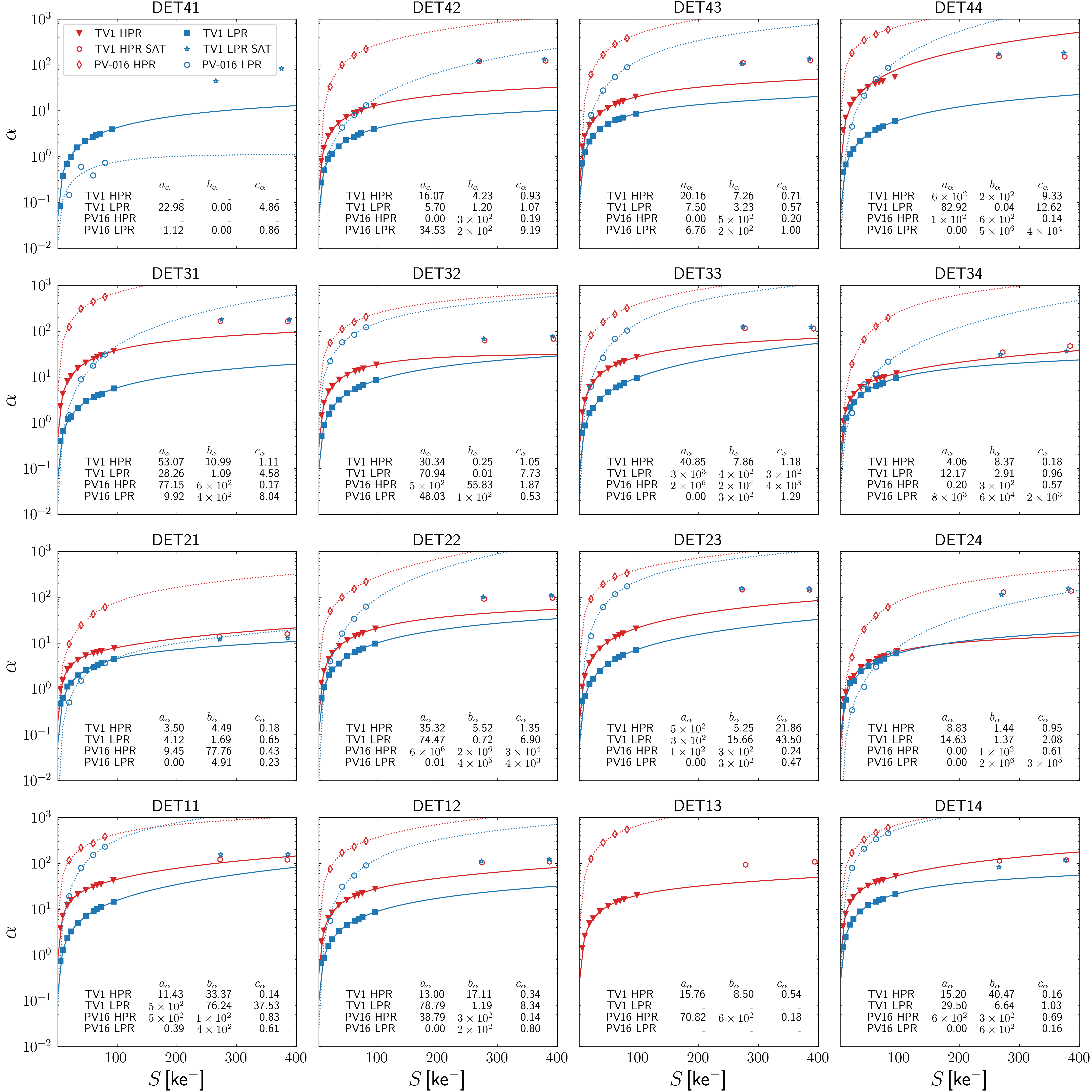}
    \caption{Persistence amplitude parameters $\alpha$ (TV1) and $\tilde{\alpha}$ (PV-016) as a function of source fluence level. Median values per detector region are reported. For the TV1 data set, measurements below and above saturation are shown separately. The solid and dotted curves represent fits of Eq.~(\ref{eq:alpha_S}) to the below-saturation TV1 and PV-016 data sets, respectively. The best-fit parameter values are reported for completeness only. For some detectors, HPR (DET41) or LPR (DET13) results are not shown because of the detector characteristics, as only one of the two regions is present.}
    \label{fig:alpha_fpa}
    \end{center}
\end{figure*}

\begin{figure*}
    \centering
    \includegraphics[width=\linewidth]{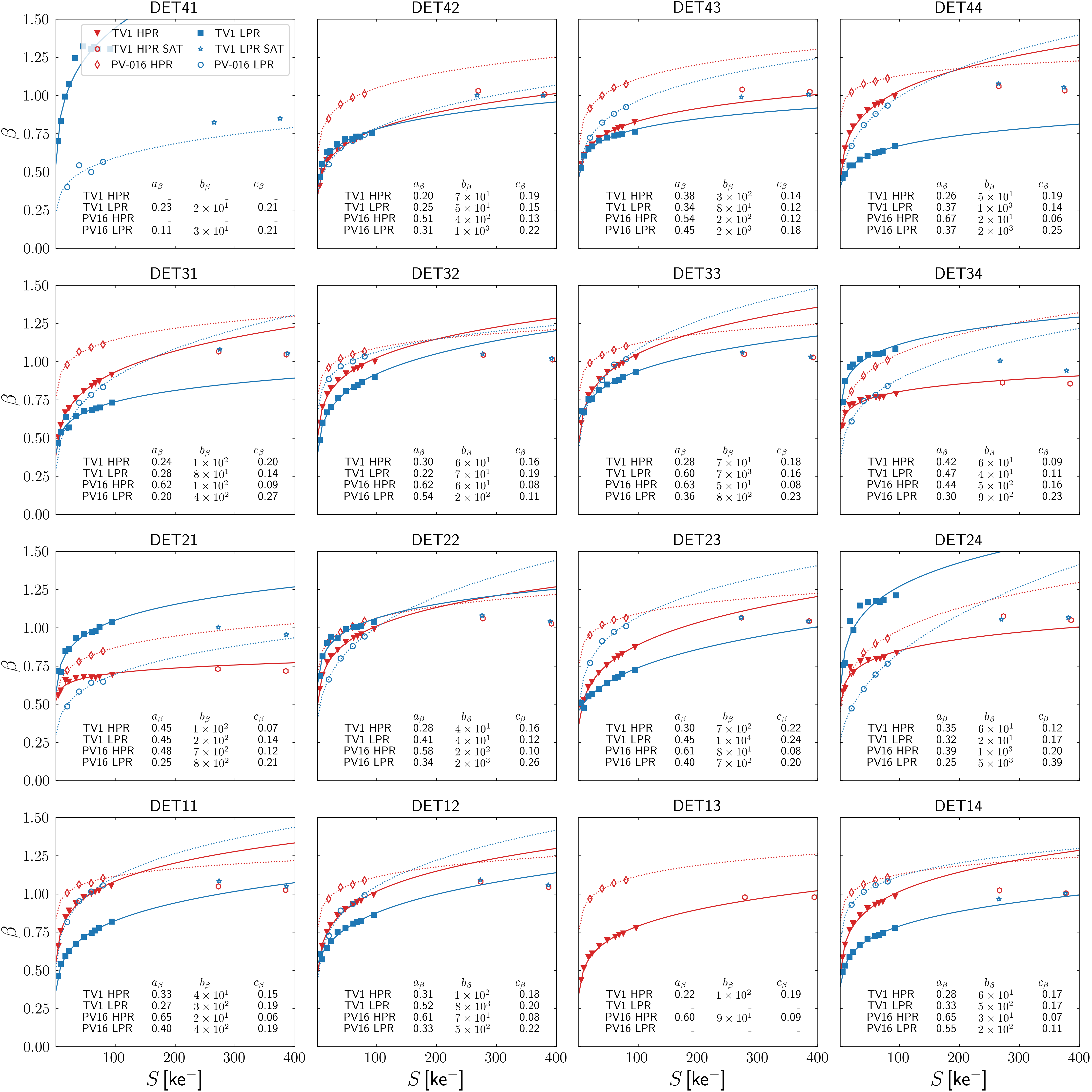}
    \caption{Persistence decay index $\beta$ (TV1) and $\tilde{\beta}$ (PV-016) as a function of source fluence level. Median values per detector region are reported. For the TV1 data set, measurements below and above saturation are shown separately. The solid and dotted curves represent fits of Eq.~(\ref{eq:beta_S}) to the below-saturation TV1 and PV-016 data sets, respectively. The best-fit parameter values are reported for completeness only. For some detectors, HPR (DET41) or LPR (DET13) results are not shown because of the detector characteristics, as only one of the two regions is present.}
    \label{fig:beta_fpa}
\end{figure*}

\begin{figure*}
    \centering
    \includegraphics[width=\linewidth]{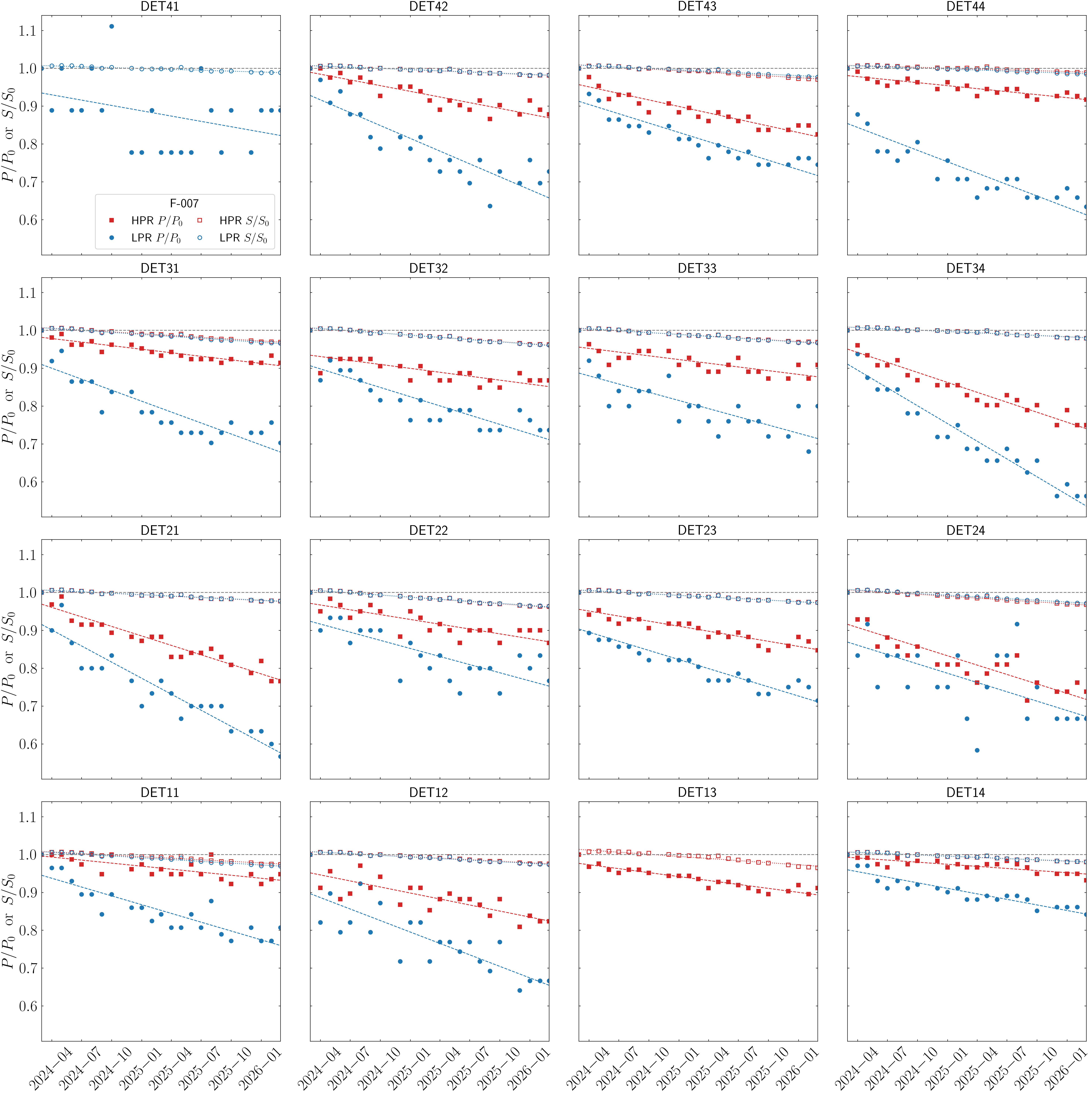}
    \caption{Temporal evolution of the persistence response in the F-007 data set, relative to the initial observations $P_0$ and $S_0$ obtained in the dark and flat-field, respectively. Here, $P_0$ and $S_0$ refer to the signals measured in April 2024. For some detectors, HPR (DET41) or LPR (DET13) results are not shown because of the detector characteristics, as only one of the two regions is present.}
    \label{fig:evolution_flight_fpa}
\end{figure*}

\begin{figure*}
    \centering
    \includegraphics[width=\linewidth]{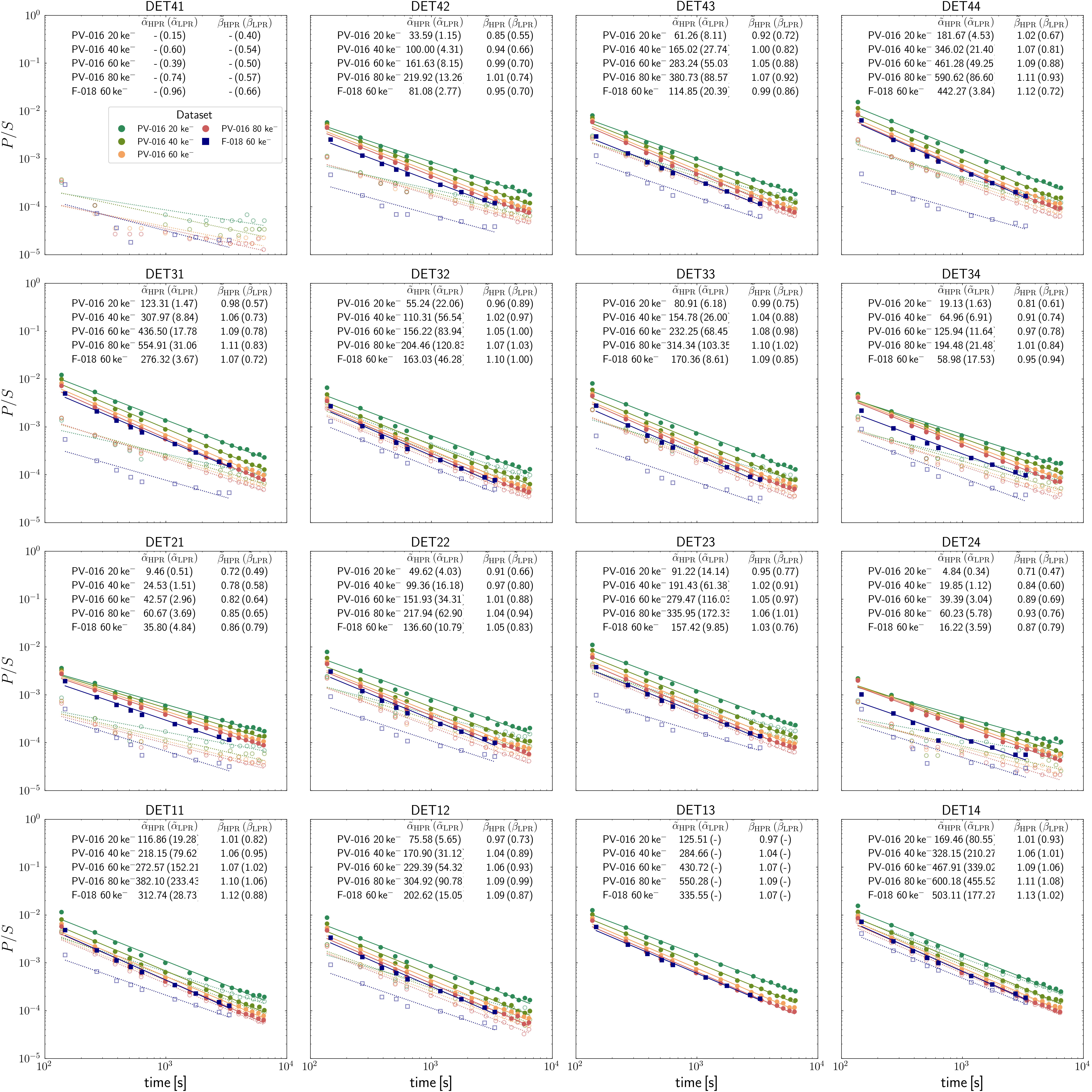}
    \caption{Persistence decay measured in the PV-016 and F-018 data sets. HPR are shown with solid lines and filled markers, LPR with dashed lines and unfilled markers. The persistence amplitudes $\tilde{\alpha}$ and the decay index $\tilde{\beta}$ are indicated in the legend for HPRs and LPRs in parentheses.  For some detectors, HPR (DET41) or LPR (DET13) results are not shown because of the detector characteristics, as only one of the two regions is present.} 
    \label{fig:decay_flight_fpa}
\end{figure*}

\begin{figure*}
    \centering
    \includegraphics[width=\linewidth]{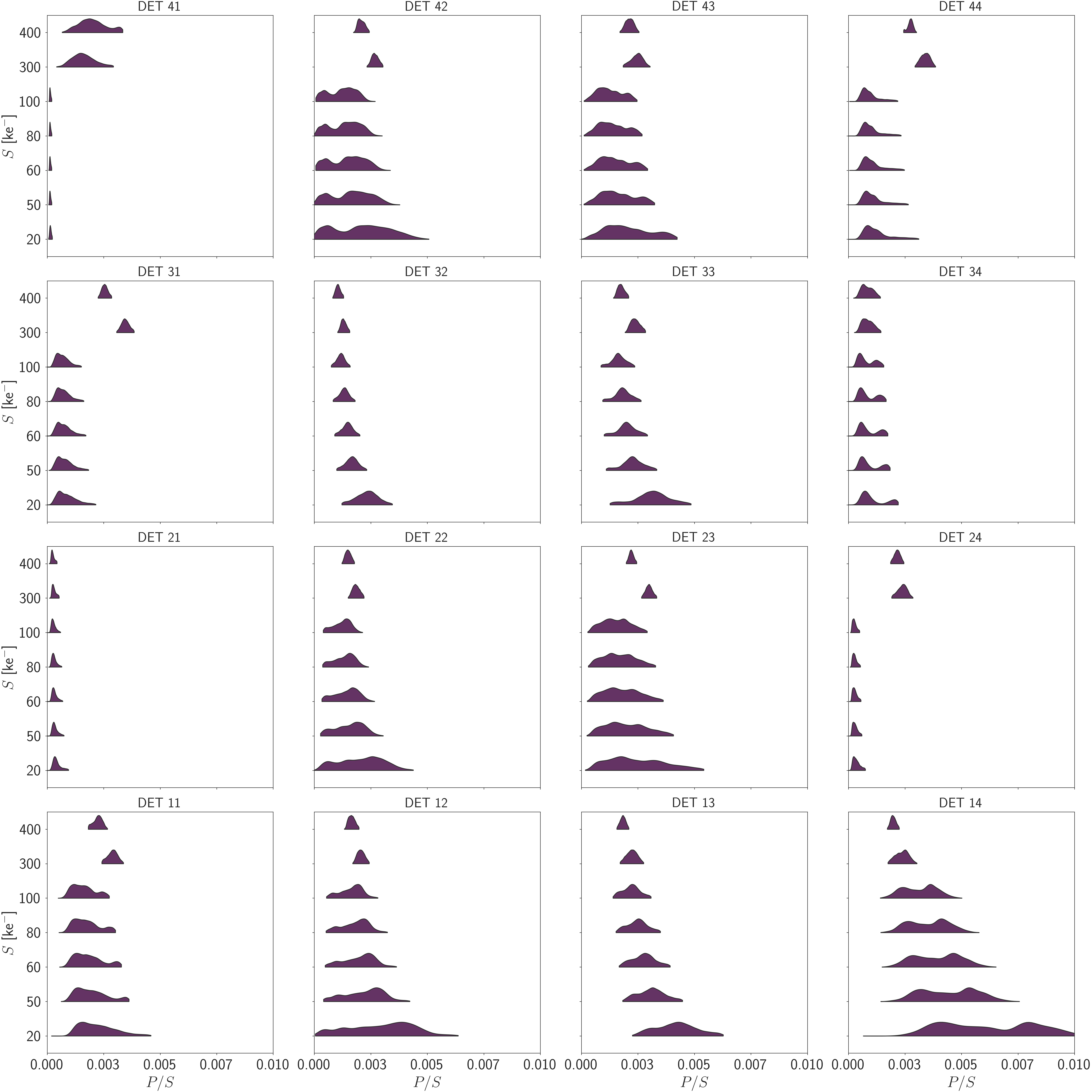}
    \caption{Evolution of persistence to fluence ration $P/S$ as a function of increasing fluence $S$ under flat-field illumination in the TV1 data set, both below and above pixel saturation (130\,ke$^-$). Each panel corresponds to a different detector and shows the distribution of persistence amplitudes for seven increasing fluence levels: five below saturation (approximately 20, 50, 60, 80, and 100\,ke$^-$) and two above saturation (approximately 300 and 400\,ke$^-$).}
    \label{fig:saturation_mosaic}
\end{figure*}

\begin{figure*}
    \centering
    \includegraphics[width=\linewidth]{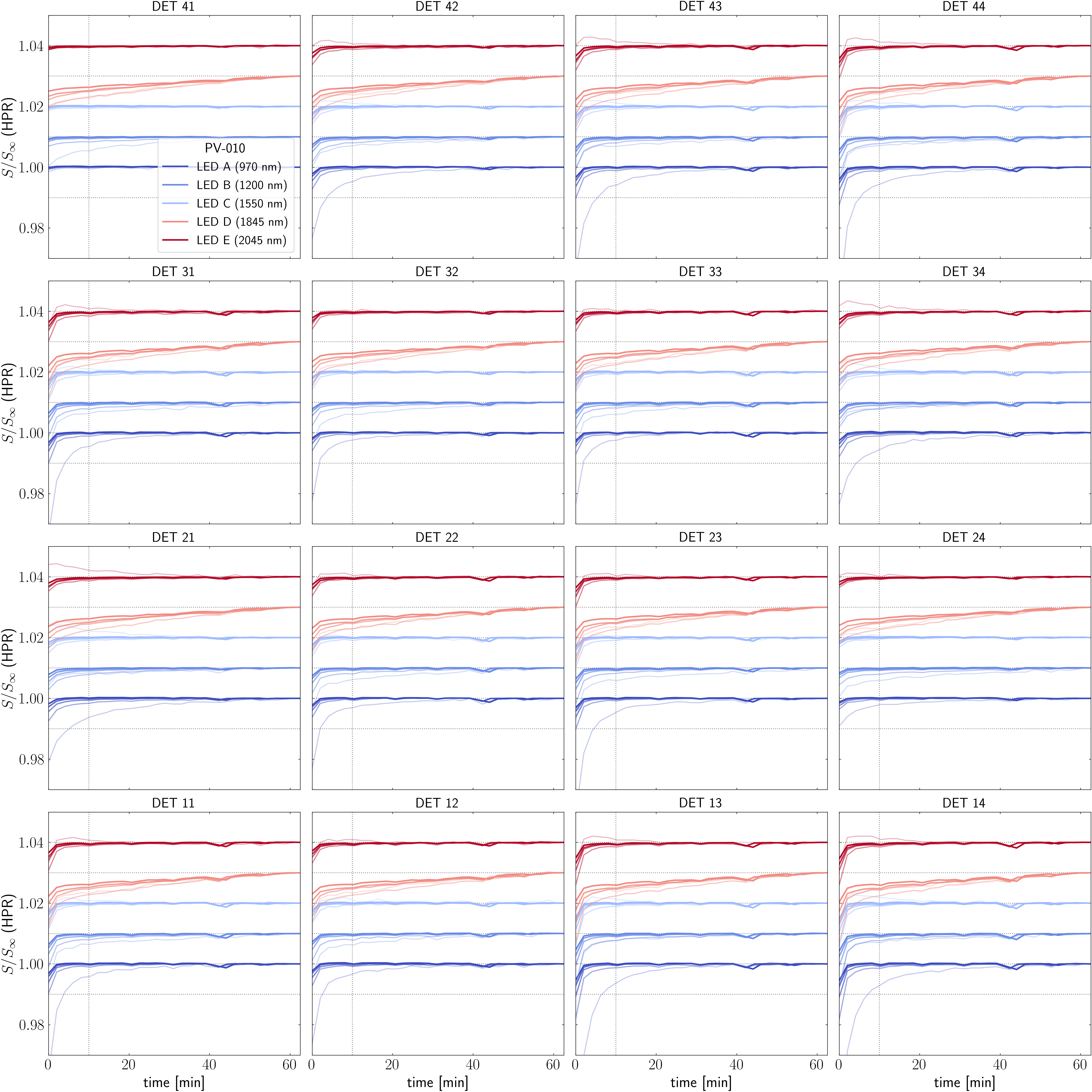}
    \caption{Impact of charge trapping and release on signal measurements. Ratio $S_i/S_\infty$ versus time for five LEDs at progressively increasing fluence levels (around 8, 18, 32, 50, and 64\,ke$^-$ per LED), with line width scaling with fluence. For clarity, the curves for LEDs B--E are vertically offset by constant amounts of +0.01, +0.02, +0.03, and +0.04, respectively. No wavelength dependence is observed. LPRs and HPRs in each detector are averaged together for clarity.} 
    \label{fig:flat_LED_time_dep}
\end{figure*}

\begin{figure*}
    \centering
    \includegraphics[width=\linewidth]{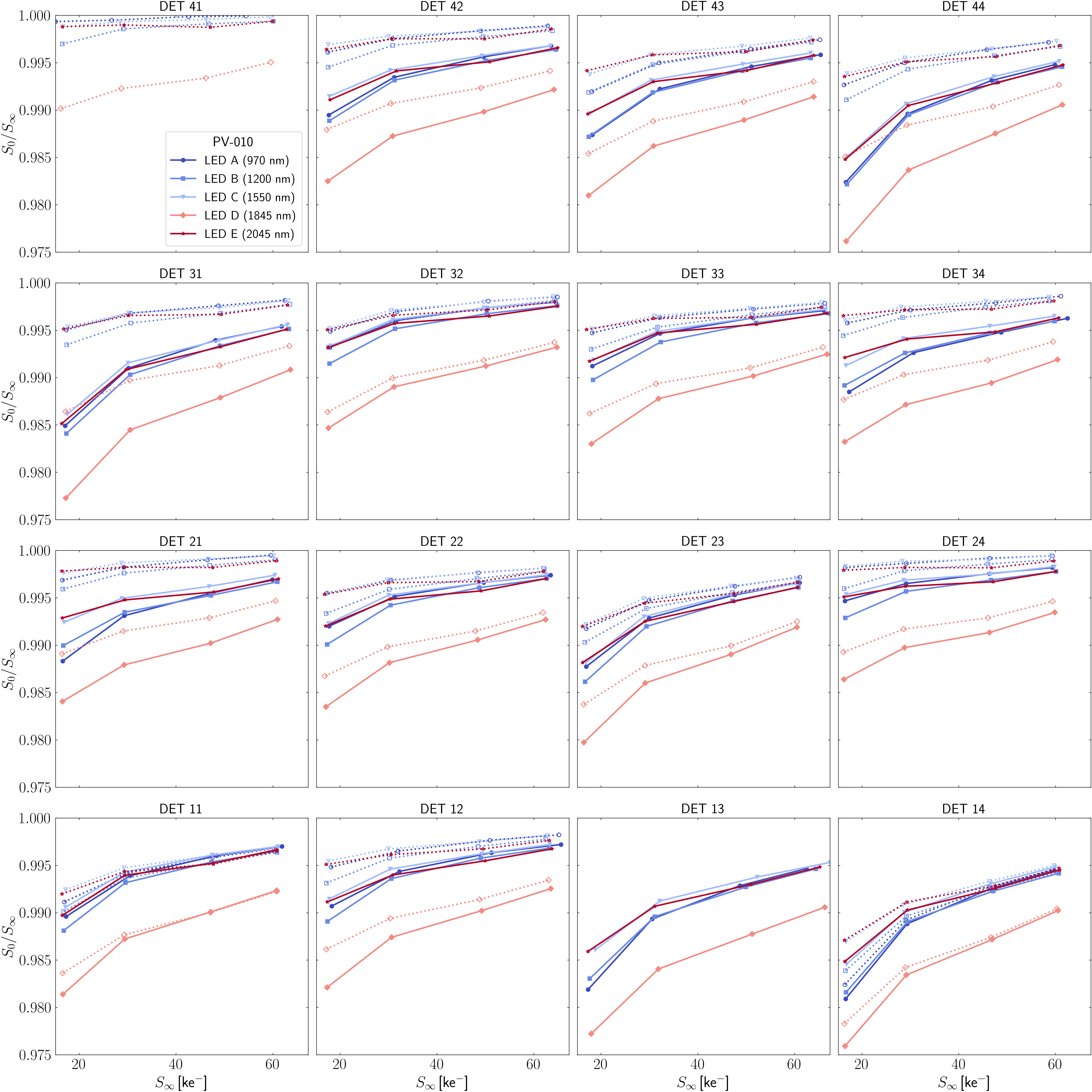}
    \caption{Ratio $S_0/S_\infty$ as function of $S_\infty$ for the five LEDs. Filled markers and solid lines correspond to HPRs. Empty markers and dotted lines correspond to LPRs.  For some detectors, HPR (DET41) or LPR (DET13) results are not shown because of the detector characteristics, as only one of the two regions is present.} 
    \label{fig:flat_LED_flux_dep}
\end{figure*}


\end{appendix}
\label{LastPage}
\end{document}